\documentclass[hyper,11pt,letterpaper]{JHEP3}
\usepackage[dvips]{epsfig}
\usepackage{amsmath,amssymb,epsf}
\usepackage{cite}
\usepackage{graphicx}
\usepackage{dcolumn}
\usepackage{bm}

\newcommand{\be}{\begin{equation}}
\newcommand{\ee}{\end{equation}}
\newcommand{\bea}{\begin{eqnarray}}
\newcommand{\eea}{\end{eqnarray}}
\newcommand{\ba}{\begin{eqnarray}}
\newcommand{\ea}{\end{eqnarray}}

\newcommand{\beq}{\begin{equation}}
\newcommand{\eeq}{\end{equation}}
\newcommand{\beqa}{\begin{eqnarray}}
\newcommand{\eeqa}{\end{eqnarray}}
\newcommand{\beqar}{\begin{eqnarray*}}
\newcommand{\eeqar}{\end{eqnarray*}}

\newcommand{\ie}{{\it i.e.}\ }

\newcommand{\N}{{\mathcal{N}}}

\def\a{\alpha}

\def\g{\gamma}

\def\lam{\lambda}
\def\m{\mu}

\def\o{\omega}  
\def\p{\pi}                

\def\L{\Lambda}
\def\O{\Omega}

\def\nc{N_c}
\def\nf{N_f}

\def\nfour{\mathcal N\,{=}\,4}
\def\ntwo{\mathcal N\,{=}\,2}
\def\none{\mathcal N\,{=}\,1}
\def\Om{{\cal{O}}_m}
\def\Omv{\langle {\cal{O}}_m \rangle}
\def\Ophi{{\cal{O}_{\phi}}}
\def\Ophiv{\langle {\cal{O}_{\phi}} \rangle}

\def\rarrow{\rightarrow}
\def\j{J^{\mu}}
\def\jtv{\langle J^t \rangle}
\def\lag{{\cal L}}

\title{\LARGE Toward a Holographic Model of Superconducting Fermions}

\author{Andy O'Bannon\footnotemark[1]\,\\
Max Planck Institut f\"{u}r Physik (Werner Heisenberg Institut) \\ F\"{o}hringer Ring 6, 80805 M\"{u}nchen, Germany}

\footnotetext[1]{E-mail: \email{ahob@mppmu.mpg.de}}

\abstract{We use the AdS/CFT correspondence to study $\N=4$ supersymmetric $SU(N_c)$ Yang-Mills theory, in the limits of large $N_c$ and large 't Hooft coupling, coupled to a number $N_f$ of massless hypermultiplet fields in the fundamental representation of the gauge group. We identify a $U(1)$ subgroup of the R-symmetry under which the fermions in the hypermultiplet are charged but the scalars are not. All the hypermultiplet fields are also charged under a $U(1)$ baryon number symmetry. We introduce an external magnetic field for the baryon number $U(1)$, which triggers the spontaneous breaking of the $U(1)$ R-symmetry, and we then introduce a chemical potential for the $U(1)$ R-charge, producing a state with a nonzero density of the $U(1)$ R-charge. The system should then exhibit superconductivity of the $U(1)$ R-charge. The dual supergravity description is a number $N_f$ of D7-branes in $AdS_5 \times S^5$ with angular momentum on the $S^5$ and a worldvolume magnetic field. We study the zero-temperature thermodynamics of the system, and find that for sufficiently large magnetic field the system prefers to be in the symmetry-broken phase. For smaller magnetic fields we find a discontinuous free energy, indicating that our gravitational setup does not capture all equilibrium states of the field theory.}

\keywords{AdS/CFT, D-branes, Brane dynamics in gauge theories}

\preprint{MPP-2008-141}

\begin{document}

\section{Introduction}

The anti- de Sitter / Conformal Field Theory (AdS/CFT) correspondence \cite{Maldacena:1997re,Gubser:1998bc,Witten:1998qj}, and more generally gauge-gravity duality, provides a novel tool for studying strongly-coupled systems. In particular, gauge-gravity duality provides tractable examples of strongly-coupled gauge theories, and thus may shed light on the low-energy dynamics of the theory of the strong interactions, Quantum Chromodynamics (QCD). More recently, gauge-gravity duality has been used to study physical phenomena familiar from solid state physics, such as the quantum Hall effect \cite{KeskiVakkuri:2008eb,Davis:2008nv} and superconductivity \cite{Hartnoll:2008vx,Gubser:2008zu,Gubser:2008wv,Horowitz:2008bn,Hartnoll:2008kx,Ammon:2008fc,Basu:2008bh}. Gauge-gravity duality may thus provide solvable toy models for condensed matter systems.

In this paper we will: 1.) identify a specific gravitational system whose field theory dual includes a global $U(1)$ symmetry under which only fermions are charged, 2.) study states of that theory with a finite density of the $U(1)$ charge, and 3.) study a mechanism that triggers the spontaneous breaking of the $U(1)$ symmetry, such that the system should be in a superconducting state. We will not study the transport properties of the system, however: in this paper we will study only zero-temperature, finite-density thermodynamics.

We will study a conformal field theory, namely $\N=4$ supersymmetric $SU(N_c)$ Yang-Mills theory (SYM), in the 't Hooft limit $N_c \rarrow \infty$ with the 't Hooft coupling $\lambda \equiv g_{YM}^2 N_c$ fixed, and in the additional limit $\lambda \gg 1$. We will introduce a number $N_f$ of $\N=2$ supersymmetric hypermultiplets that transform in the fundamental representation of the gauge group, \textit{i.e.} flavor fields. An $\N=2$ hypermultiplet contains two complex scalars and two Weyl fermions of opposite chirality. We will call the scalars ``squarks'' and the fermions ``quarks,'' in analogy with (supersymmetric) QCD. We will work in the ``probe limit,'' in which we keep $N_f$ fixed as we send $N_c \rarrow \infty$, so that $N_f \ll N_c$, and work to leading order in the small parameter $N_f/N_c$.

The AdS/CFT correspondence is the statement that $\N=4$ SYM theory in the limits above is equivalent to type IIB supergravity on the near-horizon geometry of very many D3-branes, which is $AdS_5 \times S^5$, where $AdS_5$ is (4+1)-dimensional anti- de Sitter space and $S^5$ is a five-sphere \cite{Maldacena:1997re,Gubser:1998bc,Witten:1998qj}. The correspondence is ``holographic'' in the sense that the (3+1)-dimensional field theory is dual to gravity in the (4+1)-dimensional AdS space. The $N_f$ hypermultiplets appear in the supergravity description as a number $N_f$ of D7-branes embedded in the $AdS_5 \times S^5$ geometry \cite{Karch:2002sh}.

$\N = 4$ SYM theory has an $SO(6)_R$ R-symmetry (hence the subscript). The flavor fields explicitly break this to $SO(4)_R \times U(1)_R$. Of the fields in the hypermultiplet, only the fermions are charged under the $U(1)_R$. Indeed, the $U(1)_R$ acts as a chiral symmetry, rotating left- and right-handed quarks oppositely, just like the $U(1)$ axial symmetry of QCD. We may thus introduce a chemical potential for the $U(1)_R$ and produce a state with a finite density of quarks only.

To break the $U(1)_R$ spontaneously, we will exploit the fact that the hypermultiplet fields have a second $U(1)$ symmetry. With $N_f$ flavors of mass-degenerate hypermultiplets, the SYM theory will have a global $U(N_f)$ symmetry. We may identify the overall, diagonal $U(1)$ as baryon number, $U(1)_B$. We may introduce a non-dynamical, background magnetic field, $B$, for the $U(1)_B$. Previous AdS/CFT studies have shown that such a background magnetic field causes spontaneous breaking of the $U(1)_R$ symmetry \cite{Filev:2007gb,Filev:2007qu,Albash:2007bk,Erdmenger:2007bn}. We thus want to study a state with a finite $U(1)_R$ density of quarks and a $U(1)_B$ magnetic field.

What is the gravitational description of such a state? The SYM R-symmetry is dual to isometries of the $S^5$, and a finite R-charge density in the SYM theory is dual to a state in supergravity with angular momentum in the $S^5$ directions \cite{Gubser:1998jb,Kraus:1998hv,Cvetic:1999ne,Cvetic:1999xp}. To introduce a $U(1)_R$ density of quarks, then, we must study a D7-brane rotating in the $S^5$ directions. D7- and D5-branes spinning in $AdS_5$ backgrounds have been studied in refs. \cite{Filev:2008xt,Evans:2008zs}, respectively. The $U(1)_B$ symmetry is dual to the $U(1)$ gauge invariance on the worldvolume of the D7-branes, hence to introduce a magnetic field in the SYM theory we must introduce a magnetic field on the worldvolume of the D7-branes.

Notice that the $U(1)_R$ symmetry will \textit{not} be broken due to finite density physics, rather, we will break the symmetry by an external mechanism, the $U(1)_B$ magnetic field. Contrast this with the usual physics of Bose-Einstein condensation as described by a theory of a single complex scalar field with a potential including a mass term and a quartic term. The theory has a global $U(1)$ symmetry which shifts the phase of the field. Roughly speaking, a chemical potential for the $U(1)$ acts as a negative mass-squared. If the mass-squared is positive, so that the potential has a single minimum and classically the field has zero expectation value, a sufficiently large chemical potential causes a second-order phase transition to a state of broken symmetry: the potential changes to a ``wine bottle'' shape. Our system is analogous to a scalar field with a \textit{negative} mass-squared, in the sense that the symmetry is broken even at zero chemical potential. In particular, if our system does describe a superconductor, any ``pairing'' mechanism will be due to the magnetic field, and not the due to high-density physics.

In addition, as the $U(1)_R$ is an R-symmetry, some of the fields of the $\N=4$ multiplet also carry the $U(1)_R$ charge, and hence may also contribute to a state of finite $U(1)_R$ density. We will ignore such a contribution. In supergravity language, we will only study D7-branes spinning in $AdS_5 \times S^5$, rather than D7-branes spinning in the near-horizon geometry of spinning D3-branes \cite{Gubser:1998jb,Kraus:1998hv,Cvetic:1999ne,Cvetic:1999xp}. In SYM theory language, we will introduce different values of the chemical potential for different fields: zero for the adjoint fields and nonzero for the flavor fields. We discuss the relevant background (of spinning D3-branes and $\N=4$ SYM theory with $U(1)_R$ chemical potentials \cite{Yamada:2006rx,Yamada:2008em,Hollowood:2008gp}) below, in section \ref{adjointdensity}. In particular, the $\N=4$ SYM theory (in flat space) with a $U(1)_R$ chemical potential has no genuine equilibrium ground state \cite{Yamada:2006rx,Yamada:2008em,Hollowood:2008gp}, so by ignoring the chemical potential for the adjoint fields we are ignoring a known instability of the theory.

Despite its shortcomings, we hope that our system may serve as a nice toy model, \textit{i.e.} as a laboratory for questions about gravitational descriptions of superconductivity. This system is attractive mainly because it is relatively simple and because the dual field theory is explicitly known (in particular, we know that in the flavor sector only fermions carry the $U(1)_R$ charge). Furthermore, we believe this system is more attractive than some other systems with similar features, for a number of reasons.

Other gravitational systems dual to field theories with only fermions in the flavor sector include for example the D4/D6 and D4/D8 models of refs. \cite{Kruczenski:2003uq,Sakai:2004cn}. An important difference between our system and these systems is the potential between heavy test charges in the field theory. The D4/D6 and D4/D8 systems describe field theories with confining potentials, which are of course preferable when the goal is to study QCD. The $\N=4$ SYM theory with massless $\N=2$ flavor fields in the probe limit is conformal, hence the potential between heavy test charges is necessarily Coulombic, which may be preferable for some condensed matter applications. More generally, condensed matter systems near quantum critical points are typically described by strongly-coupled conformal field theories (see ref. \cite{Herzog:2007ij} and references therein).

Other gravitational descriptions of field theories with spontaneously broken $U(1)$ chiral symmetry include for example the D4/D6 system as well as D7-branes in the Constable-Myers background \cite{Babington:2003vm}. The benefit of using the $U(1)_B$ magnetic field to break the symmetry is that, in some sense, it is intrinsic to the D7-brane, \textit{i.e.} in field theory language we can change the scale of chiral symmetry breaking (which is determined by the magnetic field $B$) without changing any other physics. In contrast, consider the D4/D6 system of ref. \cite{Kruczenski:2003uq}. In that case the scale of chiral symmetry breaking and the scale of confinement are both set by the same parameter, the Kaluza-Klein compactification scale. If we want to change the scale of chiral symmetry breaking, then, we must also change the glueball masses.

Our main results are that for sufficiently large $U(1)_B$ magnetic field the system prefers to be in a symmetry-broken phase. For smaller values of the magnetic field we find a gap in the free energy, indicating that our supergravity setup is missing something. More specifically, our ansatz for the D7-brane embedding does not seem to capture all values of the free energy.

This paper is organized as follows. In section \ref{theoryanddual}, we describe the SYM theory and its supergravity dual in more detail. In section \ref{u1r} we study D7-branes rotating in $S^5$ directions with no $U(1)_B$ magnetic field, extending the analysis of ref. \cite{Evans:2008zs}. In section \ref{u1b} we review the results of refs. \cite{Filev:2007gb,Filev:2007qu,Albash:2007bk,Erdmenger:2007bn} for non-rotating D7-branes in AdS/CFT with worldvolume magnetic fields. In section \ref{u1randu1b} we study the full problem of rotating D7-branes with worldvolume magnetic fields. We conclude in section \ref{conclusion} with some discussion and with suggestions for future research. We collect some technical details in the Appendix.

\section{The Theory and Its Supergravity Dual}
\label{theoryanddual}

\subsection{The Theory}
\label{thetheory}

We will study the maximally supersymmetric Yang-Mills theory in (3+1) dimensions, $\N=4$ SYM theory, with gauge group $SU(N_c)$. The fields of the $\nfour$ supermultiplet include the gluons, four Weyl fermions and three complex scalars. The $\N=4$ SYM theory is conformal, so that the 't Hooft coupling, $\lambda \equiv g_{YM}^2 N_c$, is a free parameter. We will take the 't Hooft limit of $N_c \rarrow \infty$ with $\lambda$ fixed, followed by the strong-coupling limit $\lambda \gg 1$.

We will also introduce a number $\nf$ of $\ntwo$ supersymmetric hypermultiplets in the fundamental representation of the gauge group, \ie flavor fields. In the language of $\none$ supersymmetry, the $\ntwo$ hypermultiplet contains two chiral multiplets of opposite chirality: two Weyl fermions of opposite chirality and two complex scalars. We will colloquially refer to the fermions as ``quarks,'' and the scalars as ``squarks,'' in loose analogy with QCD.

We will keep $\nf$ fixed as we take $\nc \rarrow \infty$, and work to leading order in the small parameter $\nf/\nc$. This is known as the probe limit. In the language of perturbation theory, we are discarding diagrams that contain quark or squark loops. More physically, we are ignoring quantum effects due to the flavor fields because such effects are parametrically suppressed by powers of $\nf/\nc$. In particular, we ignore the running of the coupling: the beta function of the theory is proportional to $\lambda^2 \frac{\nf}{\nc}$, which vanishes in our limit, for fixed $\lambda$.

The $\N=4$ SYM theory has an $SO(6)_R$ R-symmetry. The flavor fields explicitly break this to $SO(4)_R \times SO(2)_R$. We will denote the current associated with the $SO(2)_R \equiv U(1)_R$ as $\j$. What are the charges of the fields under the $U(1)_R$? A table of charge assignments appears in many places, for example in refs. \cite{Hong:2003jm,Erdmenger:2007cm}. In the $\N=4$ multiplet, one complex adjoint scalar has charge $+2$. Two of the adjoint fermions have charge $+1$, and two have charge $-1$. In the flavor sector, the squarks are neutral under the $U(1)_R$. The fermion in one $\none$ chiral multiplet has charge $+1$, while the fermion in the other chiral multiplet has charge $-1$.

The $U(1)_R$ thus mimics the $U(1)$ axial symmetry of QCD, and we will refer to it as a chiral symmetry. Notice in particular that a quark mass term will explicitly break the $U(1)_R$ symmetry. Additionally, if $\nf$ is on the order of $\nc$ then the $U(1)_R$ is anomalous, just like the $U(1)$ axial symmetry of QCD. In the probe limit, however, the anomaly is not apparent, for the same reasons that the running of the gauge coupling is not apparent: the quantum effects that produce the anomaly are parametrically suppressed by powers of $\nf / \nc$.

If the quarks are massless, so that the $U(1)_R$ is a symmetry of the Lagrangian and $J^{\mu}$ is conserved, $\partial_{\mu} J^{\mu}=0$, then we may introduce a chemical potential for the $U(1)_R$. In a state with a finite $U(1)_R$ density, any contribution that the flavor fields make to the density must come from the quarks, and not from the squarks. Notice that, as the $U(1)_R$ symmetry is an axial symmetry, a state with a net $U(1)_R$ density of quarks is a state with an excess of left-handed quarks (for example).

\subsection{The Supergravity Dual}
\label{sugradual}

We begin with type IIB string theory, where we consider an intersection of $N_c$ coincident D3-branes and $N_f$ coincident D7-branes described by the array
\begin{equation}
\begin{array}{ccccccccccc}
   & x_0 & x_1 & x_2 & x_3 & x_4 & x_5 & x_6 & x_7 & x_8 & x_9\\
\mbox{D3} & \times & \times & \times & \times & & &  &  & & \\
\mbox{D7} & \times & \times & \times & \times & \times  & \times
& \times & \times &  &   \\
\end{array}
\end{equation}
We first consider the D3-branes alone. We take the near-horizon limit of the D3-brane geometry, which gives us $AdS_5 \times S^5$, with radius of curvature $L$ given by $L^4 = 4 \pi g_s \nc \alpha'^2$, where $g_s$ is the string coupling and $\alpha'$ is the square of the string length. We take the usual limits of $N_c \rarrow \infty$ with $g_s N_c$ fixed, followed by $g_s \nc \gg 1$, so that $L^4 \gg \alpha'^2$. In particular, in the latter limit, heavy string modes decouple, and we may approximate the string theory by type IIB supergravity in $AdS_5 \times S^5$. The AdS/CFT correspondence is then the statement that supergravity on this background is equivalent to the low-energy theory on the D3-brane worldvolume, which is $\N=4$ SYM theory in the 't Hooft limit with large 't Hooft coupling.

The $\nf$ D7-branes introduce additional open string degrees of freedom, producing fields in the fundamental representation of the $SU(N_c)$ gauge group on the D3-branes' worldvolume. If we keep $N_f$ finite as we take $\nc \rarrow \infty$, so that $N_f \ll N_c$, we may neglect the D7-branes' contribution to the stress-energy tensor in the supergravity theory\footnote{Additionally, the D7-branes source the dilaton and axion, but again, in the probe limit we neglect this effect. This is dual to the SYM theory statement that when $\nf \ll \nc$ the quantum effects of the flavor fields that cause the running of the coupling and the $U(1)_R$ anomaly are suppressed.}. The D7-branes thus do not deform the geometry: they are probes embedded in $AdS_5 \times S^5$. The D7-branes will be extended in the $AdS_5$ directions as well as along an $S^3 \subset S^5$. 

We will use an $AdS_5 \times S^5$ metric suited to the symmetries of the D7-branes,
\beq
\label{adsmetriczerotemp}
ds^2 = \frac{r_6^2}{L^2} \, \eta_{\mu \nu} dx^{\mu} dx^{\nu} + \, \frac{L^2}{r_6^2} \, \left( dr^2 + r^2 ds^2_{S^3} + dy^2 + y^2 d\phi^2 \right)
\eeq
Here $r_6$ is the distance to the D3-branes in the transverse $\mathbb{R}^6$, $ds_{S^3}^2$ is the metric of a unit-radius $S^3$, and $dy^2 + y^2 d\phi^2$ is the metric of the $x_8\,$-$x_9$ plane written in polar coordinates. Notice that $r_6^2 = r^2 + y^2$. The boundary of $AdS_5$ is located at $r_6 \rarrow \infty$. Starting now, we will use units in which $L \equiv 1$ unless stated otherwise. We then translate between string theory and SYM theory quantities using $\alpha'^{-2} = 4 \pi g_s \nc = g_{YM}^2 \nc  = \lambda$.

The part of the D7-brane action that will be relevant here is the Dirac-Born-Infeld (DBI) term,
\beq
\label{originaldbi}
S_{D7} = - N_f T_{D7} \int d^8 \zeta \sqrt{-det\left(g_{ab}^{D7} + (2\pi\alpha') F_{ab} \right)}
\eeq
Here $T_{D7}$ is the D7-brane tension, $\zeta^a$ are the worldvolume coordinates, $g_{ab}^{D7}$ is the induced worldvolume metric, and $F_{ab}$ is the $U(1)$ worldvolume field strength.

The D7-brane has two worlvolume scalars, $y$ and $\phi$. An ansatz for the D7-brane scalars that preserves the Lorentz invariance of the Minkowski space directions, and the $SO(4) \times SO(2)$ isometry, is $F_{ab} = 0$, $\phi = 0$ and $y(r)$. The induced D7-brane metric is then
\beq
\label{inducedd7branemetric}
ds_{D7}^2 \, = \, r_6^2 \, \eta_{\mu \nu} dx^{\mu} dx^{\nu} + \, \frac{1}{r_6^2} \, \left( dr^2 \, (1 + y'(r)^2) \, + \, r^2 \, ds^2_{S^3} \right),
\eeq
and the D7-brane action is
\beq
\label{nospind7action}
S_{D7} = -\N \, V_{\mathbb{R}^{3,1}} \, \int dr \, r^3 \sqrt{1+y'(r)^2}
\eeq
where we have defined the constant
\beq
\N \equiv N_f T_{D7} V_{S^3} = \frac{\lam}{(2 \pi)^4} N_f N_c
\eeq
where $V_{S^3} = 2 \pi^2$ is the volume of a unit-radius $S^3$ and in the second equality we have converted to field theory quantities using $T_{D7} = \frac{\a'^{-4} g_s^{-1}}{(2\p)^7} = \frac{\lam N_c}{2^5 \p^6}$. Starting now, we will drop the factor $V_{\mathbb{R}^{3,1}}$ from eq. (\ref{nospind7action}), which represents the (infinite) volume of (3+1)-dimensional Minkowski space, and re-define $S_{D7}$ as an action density.

The equation of motion for $y(r)$ is
\beq
\label{yeom}
\partial_r \, \left ( r^3 \, \frac{y'(r)}{\sqrt{1+y'(r)^2}} \right)= 0,
\eeq
which restricts the asymptotic form of solutions to be
\beq
\label{nospinyasymptoticform}
y(r) \, = \, c_0 \, + \, \frac{c_2}{r^2} \, + \, O\left(\frac{1}{r^8}\right)
\eeq
with constant coefficients $c_0$ and $c_2$. Here $c_0$ is the asymptotic separation between the D3-branes and the D7-branes in the $x_8\,$-$x_9$ plane. A string stretched between the D3-branes and D7-branes, whose endpoint represents an excitation in the fundamental representation on the D3-brane worldvolume, will have minimum length $c_0$. We may thus identify the mass $m$ of the hypermultiplet fields as this length times the string tension: $m = \frac{c_0}{2 \pi \alpha'} = \frac{\sqrt{\lambda}}{2\pi} \, c_0$.

The field $y(r)$ is dual to an operator $\Om$ in the SYM theory given by taking $\frac{\partial}{\partial m}$ of the SYM theory Lagrangian. The operator $\Om$ thus includes the mass operator of the quarks, $m$ times the mass operator of the squarks, and a cubic coupling between the squarks and one complex scalar of the $\N=4$ multiplet (the scalar with charge $+2$ under the $U(1)_R$). $\Om$ is written explicitly in the Appendix. We also show in the Appendix that the sub-leading coefficient in eq. (\ref{nospinyasymptoticform}), $c_2$, is related to the expectation value of $\Om$ as
\beq
\Omv =  -\frac{1}{(2\pi)^3} \sqrt{\lambda} \, \nf \, \nc \, 2\, c_2
\eeq
Notice that $\Om$ is charged under the $U(1)_R$ symmetry (just as the quark mass operator is charged under the $U(1)$ axial symmetry of QCD), and hence, when $c_0=0$, a nonzero expectation value for $\Om$ signals the spontaneous breaking of the $U(1)_R$ symmetry.

The D7-brane action, eq. (\ref{nospind7action}), depends only on the derivative of $y(r)$, and hence the quantity in parentheses in eq. (\ref{yeom}) is a constant of motion. Solutions with nonzero values of the constant of motion have been studied in refs. \cite{Herzog:2007kh,Karch:2007br}. These solutions are not supersymmetric, and in fact describe D7/anti-D7 configurations. For more details, see refs. \cite{Herzog:2007kh,Karch:2007br}.

The factor under the square root in eq. (\ref{nospind7action}) is a sum of squares, so the solution with the smallest value of the on-shell action, which must be the physically preferred solution, is $y'(r) = 0$, or $y(r) = c_0$, in which case the constant of motion is zero. These embeddings are supersymmetric. In the SYM theory, they describe hypermultiplet fields with mass $m = \frac{\sqrt{\lambda}}{2 \pi} \, c_0$ and a vacuum state in which $\Omv = 0$.

In the supergravity picture, the $y(r) = c_0$ solutions describe D7-branes that ``end'' somewhere in $AdS_5$, which is most easily seen from the induced D7-brane metric evaluated on such a solution,
\beq
\label{zerospininducedd7branemetric}
ds_{D7}^2 \, = \, (r^2+c_0^2) \, \eta_{\mu \nu} dx^{\mu} dx^{\nu} + \, \frac{1}{(r^2+c_0^2)} \, \left( dr^2 + \, r^2 \, ds^2_{S^3} \right).
\eeq
At the boundary, $r \rarrow \infty$, the induced metric approaches $AdS_5 \times S^3$. When $r \rarrow 0$, however, we see that $r_6^2 = r^2 + y^2 \rarrow c_0^2$, so when $c_0$ is nonzero, the D7-brane does not reach $r_6 = 0$, and we say that the D7-brane ends at $r_6 = c_0$. From the induced metric eq. (\ref{zerospininducedd7branemetric}), we can see that when $r=0$, the $S^3$ has zero volume. What is happening as $r$ decreases is that the $S^3 \subset S^5$ ``slips'' or ''shrinks,'' as allowed by topology, and eventually collapses to a point at $r = 0$. Notice that if $c_0 = 0$ (dual to massless hypermultiplets), then the $S^3$ does not slip, and the D7-brane is present at all values of $r_6$.

The Ricci scalar $R(r)$ associated with the D7-brane's induced metric in eq. (\ref{zerospininducedd7branemetric}) is
\beq
R(r) = - \, \frac{8 \, c_0^2 + 14 \, r^2}{c_0^2 + r^2}
\eeq
At the boundary $r \rarrow \infty$, $R(r) \rarrow -14$, which is indeed the curvature of $AdS_5 \times S^3$. At the endpoint $r \rarrow 0$, $R(r) \rarrow -8$. In later sections we will compute the Ricci scalar of the D7-brane numerically, and these limits will provide useful checks.

In subsequent sections we will encounter embeddings for which the curvature diverges at $r=0$. For a general solution $y(r)$, an easy way to understand such a singularity is by looking at the induced metric in eq. (\ref{inducedd7branemetric}). To avoid an angular deficit, and hence a conical singularity, at $r=0$, we must have $(1+y'(r)^2) \rarrow 1$, hence $y'(r) \rarrow 0$. Solutions for which $y'(0)$ is nonzero are thus singular, as we will see below. When the curvature grows we expect curvature corrections to the DBI action eq. (\ref{originaldbi}) to become important, therefore we must discard such embeddings as unphysical: they are solutions to an equation of motion that arises from an action that is no longer a reliable approximation to the actual D7-brane action.

We want to study a state of the SYM theory with a nonzero $U(1)_R$ density. The global $U(1)_R$ symmetry of the SYM theory is dual to the $SO(2)$ isometry that rotates $x_8$ and $x_9$ into one another, or equivalently that shifts $\phi$ by a constant. Notice for instance that a finite $c_0$ explicitly breaks the $SO(2)$ isometry in the $x_8$-$x_9$ plane, which is dual to the statement in the SYM theory that a finite mass $m$ explicitly breaks the $U(1)_R$ symmetry. A state of the SYM theory with finite $U(1)_R$ charge density is dual to a supergravity solution with nonzero angular momentum in $\phi$. We thus want to study a D7-brane spinning in the $x_8\,$-$x_9$ plane, with a non-trivial $y(r)$: if $y(r) = 0$, so that the D7-brane sits at the origin of the $x_8\,$-$x_9$ plane for all $r$, then obviously the D7-brane will have zero angular momentum in $\phi$. Moreover, in the SYM theory we want massless hypermultiplet fields, so that the $U(1)_R$ is a symmetry of the Lagrangian, hence we want solutions with $c_0=0$. For such solutions, the leading term in the asymptotic form of $y(r)$ will be the $c_2/r^2$ term.

To summarize: our goal is to find embeddings in which the D7-brane rotates in $\phi$ and has zero $c_0$ with nonzero $c_2$, \textit{i.e.} the D7-brane has zero asymptotic separation from the D3-branes but nonzero angular momentum in the $x_8\,$-$x_9$ plane.

Notice what supergravity is telling us about the SYM theory: for a D7-brane with $c_0=0$ to have nonzero angular momentum, $c_2$ must be nonzero. Translating to SYM theory language, in any state with a finite $U(1)_R$ density, the $U(1)_R$ symmetry must be spontaneously broken, as indicated by a nonzero $\Omv$.

In AdS/CFT, we identify the supergravity action, when evaluated on a solution, with the generating functional of the SYM theory \cite{Maldacena:1997re,Gubser:1998bc,Witten:1998qj}. More specifically, with a nonzero chemical potential we identify the on-shell supergravity action $S_{D7}$ with the thermodynamic potential in the grand canonical ensemble, $\Omega$, as $S_{D7} = - \Omega$. For spinning D-branes, we identify the angular frequency of rotation, $\o$, with the chemical potential, $\m$, and the angular momentum with the density, $\jtv$ \cite{Gubser:1998jb,Kraus:1998hv,Cvetic:1999ne,Cvetic:1999xp}. In SYM theory language, the latter is given by $\jtv  = - \frac{d \Omega}{d \mu}$, so in supergravity language, we have $\jtv = \frac{d S_{D7}}{d \o}$. Notice that we will thus be studying densities proportional to the factor $\N$ in $S_{D7}$, that is, we will be studying densities of order $\jtv \propto \lambda \nf \nc$. We write an explicit formula for $\frac{d S_{D7}}{d \o}$ in the Appendix.

Embeddings for probe D5-branes spinning in $AdS_5 \times S^5$ were studied in ref. \cite{Evans:2008zs}. In that case, the dual SYM theory includes flavor fields confined to a (2+1)-dimensional defect. The Lagrangian of this theory is written explicitly in refs. \cite{DeWolfe:2001pq,Erdmenger:2002ex}. Again, in that theory, of the fields in the fundamental representation, only the fermions carry the relevant $U(1)_R$ charge. Our results for the D7-brane in $AdS_5 \times S^5$ in section \ref{u1r} will be similar to those for the D5-brane. Our results in section \ref{u1randu1b} for the D7-brane with a worldvolume magnetic field will be new.

\subsection{The Adjoint Contribution to the Density}
\label{adjointdensity}

As mentioned in section \ref{thetheory}, one of the complex scalars and all of the fermions of the $\N=4$ multiplet are charged under the $U(1)_R$, hence these fields may contribute to a state with a finite $U(1)_R$ density. A correct thermodynamic analysis must include all microstates that produce the same macroscopic charge density, hence a correct thermodynamic analysis must include states in which the fields of the $\N=4$ multiplet contribute to the density. In other words, in the grand canonical ensemble we choose values of the temperature and chemical potential, and the dynamics of the theory then determines the ground state.

In supergravity language, including the adjoint fields means allowing the D3-branes to spin in the $x_8\,$-$x_9$ plane. The near-horizon geometry of spinning D3-branes is known \cite{Gubser:1998jb,Kraus:1998hv,Cvetic:1999ne,Cvetic:1999xp}. A complete supergravity analysis would thus involve allowing D7-branes to spin in the background produced by spinning D3-branes. Notice, however, that we may give the D3-branes and D7-branes distinct angular frequencies. If the frequencies are equal, then in the SYM theory we have a single chemical potential for the $U(1)_R$. If the frequencies are distinct, we have different values of the chemical potential for different fields in the theory, one value for the fields of the $\N=4$ multiplet and another value for the fields of the $\N=2$ hypermutliplet. Translating the SYM theory thermodynamic analysis to supergravity language: a correct analysis would mean studying a system of D3-branes and D7-branes spinning with the same angular frequency and finding the solution that extremizes the on-shell supergravity action.

As done in ref. \cite{Evans:2008zs}, however, we will simply ignore the rotation of the D3-branes. In SYM theory language, we will introduce a nonzero $U(1)_R$ chemical potential for the flavor fields only. From a SYM theory point of view, then, what we will do is artificial: we introduce a $U(1)_R$ chemical potential only for the quarks, and then use the $U(1)_B$ magnetic field to force them to pair and thus break the $U(1)_R$. Nevertheless, as mentioned in the introduction, we hope that this system may serve as a toy model for answering questions about holographic superconductors.

To place our analysis in context, and to understand what ignoring the $U(1)_R$ chemical potential for the adjoint fields really means, we will now briefly review the results of refs. \cite{Yamada:2006rx,Yamada:2008em,Hollowood:2008gp}, where  the $\N=4$ SYM theory in the presence of a $U(1)_R$ chemical potential was studied at both weak and strong coupling\footnote{Notice that refs. \cite{Yamada:2006rx,Yamada:2008em,Hollowood:2008gp} focus primarily on the $\N=4$ SYM theory formlated on a spatial three-sphere, so that the adjoint scalars acquire a curvature coupling that acts as a positive mass-squared. The phase structure then becomes more interesting than for the theory in flat space: for the details, see the phase diagrams in the references. Roughly speaking, we can obtain the phase diagram for the theory in flat space by taking a ``large volume'' limit in which the radius of the three-sphere goes to infinity (relative to all other scales).}. One result of refs. \cite{Yamada:2006rx,Yamada:2008em,Hollowood:2008gp} was that, in fact, for any finite $U(1)_R$ chemical potential the SYM theory has no equilibrium ground state.

First consider the $\N=4$ SYM theory in the large-$N_c$ limit, in flat space, at zero temperature, and at zero 't Hooft coupling. A $U(1)_R$ chemical potential will act as a negative mass-squared for the scalar charged under the $U(1)_R$. In the presence of a $U(1)_R$ chemical potential, then, the potential has no minimum; the theory has no equilibrium ground state. At finite coupling, the superpotential has a moduli space parameterized by mutually-commuting constant background values for the adjoint scalars, and indeed, the zero-temperature behavior persists from weak to strong coupling \cite{Yamada:2008em,Hollowood:2008gp}, \textit{i.e.} the theory at zero temperature has no equilibrium ground state.

At finite temperature, a weak-coupling analysis of the $\N=4$ SYM theory with a $U(1)_R$ chemical potential has been performed in refs. \cite{Yamada:2006rx,Hollowood:2008gp}. The principal result was that for chemical potentials below a critical value $\mu_{crit} = \sqrt{\lambda} T$, the origin of the moduli space is meta-stable. More precisely, when $\mu < \mu_{crit}$, the potential exhibits runaway behavior for large values of the scalars, but the origin of the moduli space is a local minimum with a lifetime that grows exponentially with $\nc$. The meta-stability was discovered by computing a one-loop effective potential for the scalars, plotted against the expectation values of the scalar eigenvalues. The potential barrier between the meta-stable state and the unstable state is lowest in the case of a single eigenvalue splitting from the rest. For $\mu > \mu_{crit}$, the potential barrier disappears, and with it the meta-stable state.

The finite-temperature story is qualitatively the same at strong coupling, where the system can be analyzed using AdS/CFT. In particular, in ref. \cite{Yamada:2008em}, an analysis of probe D3-branes spinning in the near-horizon geometry of spinning D3-branes revealed that the meta-stability persists to strong coupling: roughly speaking, the stack of $\nc$ spinning D3-branes ``spits out'' individual D3-branes one at a time. The supergravity picture thus nicely agrees with the field theory picture of a single eigenvalue separating from the rest and penetrating the potential barrier.

To return to our system: introducing D7-branes in the probe limit will not alter the physics of the background produced by the spinning D3-branes since in the probe limit we ignore the back-reaction of the D7-branes. If we included the rotation of the D3-branes, then, we know \textit{a priori} that the system is either meta-stable or unstable: it cannot be the ground state because the system has no ground state. In other words, if we did include the rotation of the D3-branes, then we would expect the D3-branes to carry most of the angular momentum, and indeed to exhibit runaway behavior\footnote{We will mention in passing that we can ``fix'' the instability: in SYM theory language, we can compactify the spatial directions into a three-sphere, which, for the $\N=4$ SYM theory alone, stabilizes the theory for sufficiently small chemical potential.}.

D7-brane probes in the near-horizon geometry of spinning D3-branes have been studied in refs. \cite{AlbashRcharge,Filev:2008xt}. The principal result was that the nonzero $U(1)_R$ chemical potential does not trigger spontaneous breaking of the $U(1)_R$ symmetry\footnote{Notice that the result of ref. \cite{AlbashRcharge} seems to be the opposite: that the $U(1)_R$ chemical potential causes breaking of the $U(1)_R$ in the flavor sector. As indicated in refs. \cite{Filev:2008xt,Evans:2008zs}, however, that conclusion came from using unphysical D7-brane embeddings. In fact, a $U(1)_R$ chemical potential does not cause breaking of the $U(1)_R$.}. The case of D7-branes with worldvolume magnetic fields probing the near-horizon geometry of spinning D3-branes remains to be studied.

With the above background in mind (and in particular, remembering what we are ignoring), we now turn to our analysis of D7-branes spinning in $AdS_5 \times S^5$.

\section{Finite $U(1)_R$ Chemical Potential}
\label{u1r}

To study D7-branes spinning in $AdS_5 \times S^5$, we consider the following ansatz for the D7-brane worldvolume scalars: $y(r)$ and $\phi(t,r) = \o t + f(r)$. We will thus have a D7-brane spinning with frequency $\o$ in the $x_8$-$x_9$ plane. As explained in ref. \cite{Evans:2008zs}, and similar to the system in ref. \cite{Karch:2007pd}, the $r$ dependence in $\phi(t,r)$ is required to guarantee the reality of the D7-brane action for all values of $r$, for certain embeddings.

We can also motivate the $r$ dependence in $\phi(t,r)$ via T-duality \cite{Filev:2008xt}. If we perform a T-duality in the $\phi$ direction, the D7-brane becomes a D8-brane and $\phi(t,r) \rightarrow A_{\phi}(t,r)$, hence the D8-brane now has a constant electric field pointing in the $\phi$ direction: $F_{t\phi} = \o$. From previous experience with electric fields on D-branes in AdS/CFT \cite{Karch:2007pd}, we expect that, to guarantee reality of the D7-brane action for all values of $r$, $A_{\phi}$ must have radial dependence of the form $A_{\phi}(t,r) = \o t + f(r)$ and hence, T-dualizing back to the D7-brane, we find the $\phi(t,r)$ written above.

With our ansatz, the induced metric $g_{ab}^{D7}$ of the D7-brane has components
\beq
\label{nonzerospininducedmetric}
g_{rr}^{D7} = g_{rr} + g_{yy} y'^2 + g_{\phi \phi} \phi'^2, \qquad g_{tt}^{D7} = g_{tt} + g_{\phi \phi} \dot{\phi}^2, \qquad g_{rt}^{D7} = g_{\phi \phi} \phi' \dot{\phi}
\eeq
where primes denote differentiation with respect to $r$ and dots denote differentiation with respect to $t$, and with all other components identical to eq. (\ref{inducedd7branemetric}). The D7-brane action becomes
\beq
\label{spinningd7action}
S_{D7} = - \N \int_0^{\infty} dr \, r^3 \, \sqrt{\left( 1 + y'^2 \right) \left( 1 - \dot{\phi}^2 \frac{y^2}{(y^2 + r^2)^2} \right) + y^2 \phi'^2}
\eeq
We will define the Lagrangian $\lag$ via $S_{D7} \, = - \, \int dr \, \lag$ (notice the sign). The action depends only on $\phi'(r)$, so the system has a constant of motion, which we call $c$,
\beq
\label{spinningconstantofmotion}
\frac{\delta \lag}{\delta \phi'(r)} = \N r^3 \frac{y^2 \phi'}{\sqrt{\left( 1 + y'^2 \right) \left( 1 - \o^2 \frac{y^2}{(y^2 + r^2)^2} \right) + y^2 \phi'^2}} \equiv c.
\eeq
We then solve algebraically for $\phi'(r)$,
\beq
\label{phisolzerob}
\phi'(r) = \frac{c}{y} \,  \sqrt{ \frac{\left( 1 + y'^2 \right) \left( 1 - \o^2 \frac{y^2}{(y^2 + r^2)^2} \right)}{\N^2 y^2 r^6 - c^2 } }.
\eeq
Plugging this into the action, we find
\beq
\label{spinningd7onshellaction}
S_{D7} \, = \, - \N \, \int dr \, r^3 \, \sqrt{1+y'^2} \sqrt{\frac{1 - \o^2 \frac{y^2}{(y^2 + r^2)^2}}{1 - \frac{c^2}{\N^2}\frac{1}{y^2 \, r^6}}}.
\eeq
We can derive the equation of motion for $y(r)$ either by varying the action in eq. (\ref{spinningd7action}) and then inserting the solution for $\phi'(r)$, or by varying the Legendre-transformed action $\hat{S}_{D7}$,
\bea
\label{zeroblegendre}
\hat{S}_{D7} & = & S_{D7} - \int dr \, \phi'(r) \, \frac{\delta S_{D7}}{\delta \phi'(r)} \nonumber \\ & = & -\N \int dr \, r^3 \sqrt{1+y'^2} \, \sqrt{\left( 1 - \o^2 \frac{y^2}{(y^2 + r^2)^2} \right) \, \left( 1 - \frac{c^2}{\N^2} \frac{1}{y^2 r^6} \right)}.
\eea
We will not write the equation of motion explicitly.

The numerator and denominator under the square root in the action eq. (\ref{spinningd7onshellaction}) can change sign as $r$ goes from infinity to zero. If one changes sign while the other does not, then the action will become imaginary. Both must change sign simultaneously for the action to remain real, \ie the numerator and denominator under the square root in eq. (\ref{spinningd7onshellaction}) must share a common zero. We thus find two curves in the $(r,y)$ plane, and the D7-brane must either cross both simultaneously or cross neither for the action to remain real. The first curve is (here we restore factors of the AdS radius $L$)
\beq
1-\o^2 \frac{L^4 y^2}{(y^2 + r^2)^2} = 0
\eeq
Which is just the equation for a semicircle of radius $\frac{1}{2} \o L^2$, centered at $(0,\frac{1}{2} \o L^2)$:
\beq
\label{semicircle}
\left(y - \frac{1}{2} \o L^2\right)^2 + \, r^2 \, = \, \frac{1}{4} \, \o^2 L^4.
\eeq
As noted in ref. \cite{Evans:2008zs}, the entire D7-brane spins with constant angular velocity $\o$, and its linear velocity $\o \, y(r)$ depends on $r$, as does the local speed of light, which decreases as $r$ decreases. The semicircle is where the D7-brane's linear velocity equals the local speed of light. The second curve is a cubic,
\beq
y(r) = \frac{c}{\N} \, \frac{1}{r^3}.
\eeq
Notice that the semicircle is determined only by the value of $\o$, so once we choose $\o$ it is the same for all solutions. The value of $c$, however, varies from solution to solution. For example, we will show shortly that some solutions reach $r=0$ without ever crossing the semicircle, so the numerator under the square root in eq. (\ref{spinningd7onshellaction}) remains positive for all $r$. For the action to remain real, the denominator under the square root must also be positive for all $r$, which is only possible if $c=0$. For solutions that do cross the semicircle, the value of $c$ is fixed entirely by the position where the D7-brane crosses, that is, if $(r_0,y_0)$ is the point where the D7-brane crosses the semicircle (so that $r_0$ and $y_0$ obey eq. (\ref{semicircle})), then $c = \N \, y_0 \, r_0^3$. In other words, every solution that crosses the semicircle has its own cubic curve. The qualitative behavior of $c$ as a function of the $y$ position on the semicircle is depicted in fig. \ref{c}. The maximum value of $c$ occurs at $y = \frac{5}{8} \, \o L^2$, and $c$ goes to zero at $y=0$ and $y = \o L^2$.

\FIGURE{
\includegraphics[width=0.45\textwidth]{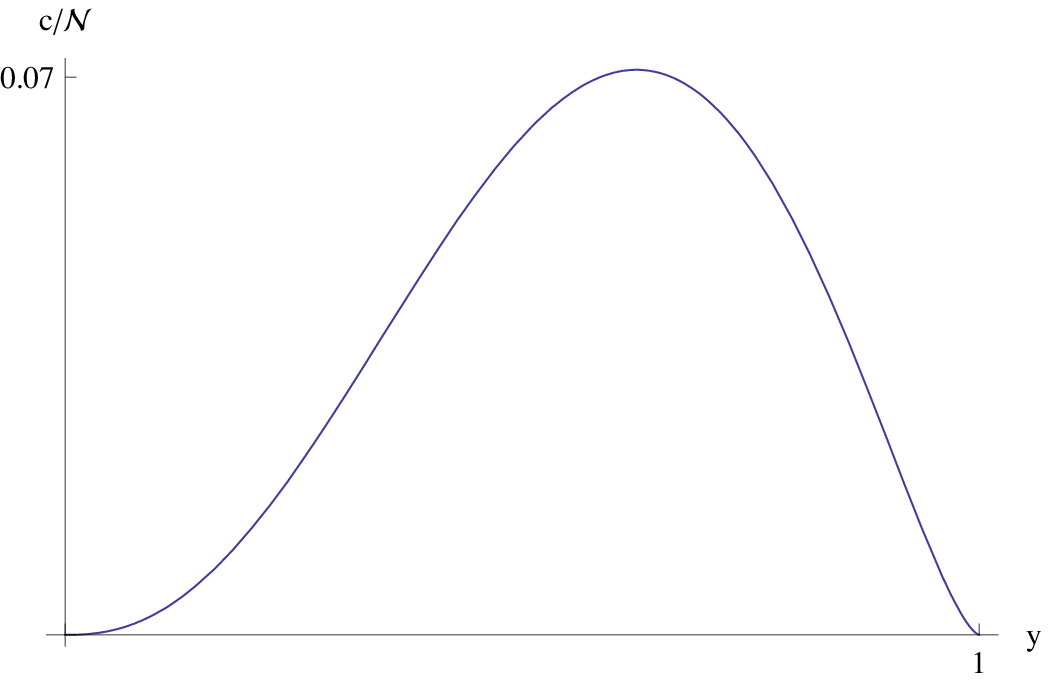}
\caption{\label{c} The value of $c/\N$ as a function of the $y$-position on the semicircle of eq. (\ref{semicircle}). Here we have chosen $\o = 1$ as an example.}}

From the equation of motion, we find the asymptotic form of $y(r)$,
\beq
\label{spinningyasymptotic}
y(r) \, = \, c_0 \, + \, \frac{c_2}{r^2} \, + \, \frac{1}{2} \, \o^2 \, c_0 \, \frac{\log r}{r^2} \, + \, O\left( \frac{\log r}{r^4} \right) 
\eeq
Notice that when $c_0$ is nonzero, a finite $\o$ produces a logarithmic term at order $1/r^2$. As mentioned above, we are interested in solutions with $c_0 = 0$, for which the logarithmic term will be absent. We show in the Appendix that, given a solution for $y(r)$ with nonzero $\o$, the expectation value of $\Om$ is determined by $y(r)$'s asymptotic coefficients as
\beq
\label{spinningd7omv}
\Omv = - \frac{1}{(2\pi)^3} \, \sqrt{\lambda} \, \nf \, \nc \, \left( \, 2 \, c_2 \, + \, \frac{1}{2} \, \o^2 \, c_0 \, + \frac{1}{2} \,  \o^2 \, c_0 \, \log (c_0^2)  \right).
\eeq
From the explicit solution for $\phi'(r)$, we can also find $\phi(t,r)$'s asymptotic form,
\beq
\label{spinningphiasymptotic}
\phi(t,r) \, = \, \o t \, - \, \frac{1}{2} \, \frac{c}{c_0^2} \, \frac{1}{r^2} \, + \, O\left( \frac{\log r}{r^4} \right)
\eeq
Notice the factor of $c_0^2$ in the denominator of the coefficient of the $1/r^2$ term, which suggests that solutions with $c_0 = 0$ (the ones we want) must have $c=0$, as otherwise $\phi(t,r)$ diverges asymptotically. Indeed, we have found numerically that this is always the case.

The field $\phi$ is dual to a SYM theory operator that we will denote $\Ophi$, which is the phase of the hypermultiplet mass operator $\Om$. We write $\Ophi$ explicitly in the Appendix.  We also show in the Appendix that the constant $c$ determines the expectation value of $\Ophi$ as $\Ophiv = c$.

Our numerical solutions for $y(r)$ are depicted in fig. \ref{ysolzerob}. We generate these as follows. We divide solutions into two classes, those that reach $r=0$ ``above the semicircle,'' and those that intersect the semicircle.

\FIGURE{
\includegraphics[width=0.49\textwidth]{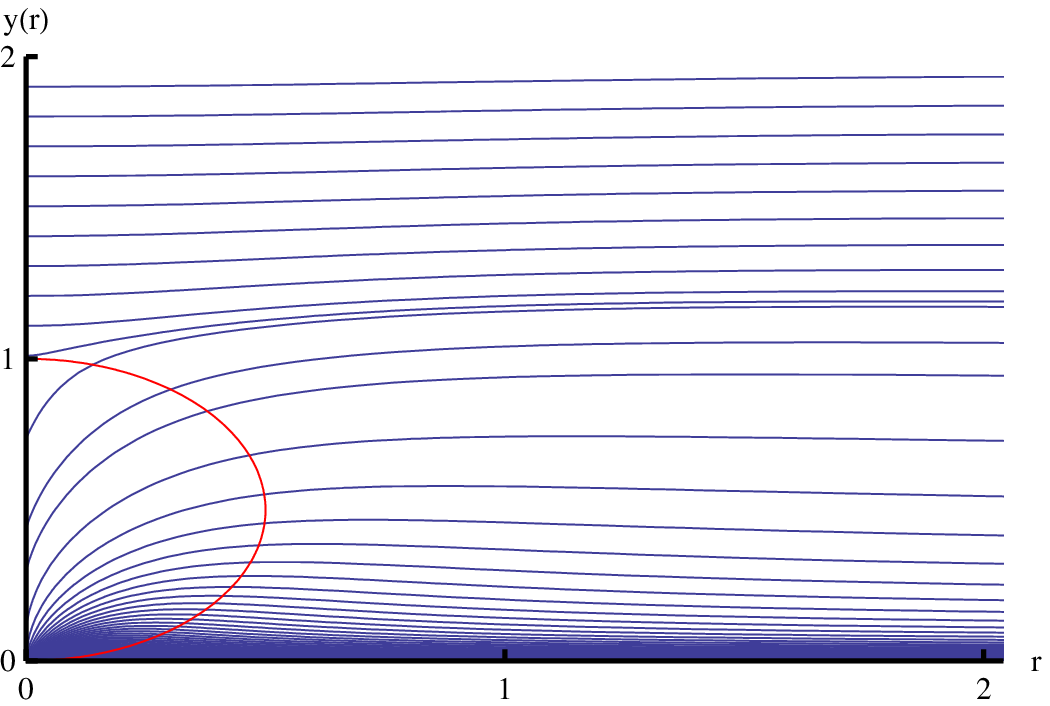}
\caption{\label{ysolzerob} Numerically-generated solutions for spinning D7-brane embedding functions $y(r)$. The semicircle of eq. (\ref{semicircle}) is also depicted. Here we have chosen $\o=1$ as an example.}}

Solutions that end above the semicircle have $c=0$. These solutions describe D7-branes for which the $S^3$ collapses to zero volume before intersecting the semicircle. To generate these, we specify the value of $y(r)$ at $r=0$, which must be $\geq \o$, and we require that $y'(r)$ vanish at $r=0$, to avoid a conical singularity, as explained in section \ref{sugradual}.

Solutions that intersect the semicircle have nonzero $c$. The equation of motion depends on $c$, so to generate solutions we first need to fix $c$, which we do simply by choosing a point $(r_0,y_0)$ on the semicircle. We then need the value of the derivative at the semicircle, $y'(r_0)$. As in ref. \cite{Evans:2008zs}, we can derive a regularity condition on $y'(r)$ from the equation of motion expanded about the semicircle. We have found numerically, however, that the solutions are amost entirely insensitive\footnote{In more detail: despite changing the value of $y'(r)$ at the semicircle by five orders of magnitude, and even changing its sign, the solutions always ``settle down'' to the solutions depicted in fig. \ref{ysolzerob} within a very short distance from the semicircle.} to the value of $y'(r_0)$. Given $(r_0,y_0)$ and $y'(r_0)$, we can numerically solve the equation of motion for all $r$.

The solutions in fig. \ref{ysolzerob} are qualitatively similar to the solutions found for spinning D5-branes in ref. \cite{Evans:2008zs}. The solutions are also qualitatively similar to the solutions found in refs. \cite{Erdmenger:2007bn,Albash:2007bq} for D7-branes with a worldvolume electric field, dual to the SYM theory in the presence of external $U(1)_B$ electric field, which is easy to understand via T-duality arguments such as the one we gave above. We will make three comments about the solutions.

First, as found in ref. \cite{Evans:2008zs} for spinning D5-branes, the only solution with $c_0 = 0$ is the trivial solution, $y(r)=0$, which has no angular momentum. In other words, using SYM theory language, for massless quarks, introducing the $U(1)_R$ chemical potential does not break the $U(1)_R$, but also does not produce a $U(1)_R$ density. We have not found a good SYM theory argument for why this is so. Notice also that all the solutions that cross the semicircle have nonzero $c_0$.

Second, all the solutions with nonzero $c_0$ describe flavor fields in the SYM theory with time-dependent masses. More specifically, as $\phi$ corresponds to $\Ophi$, the phase of the mass operator $\Om$, these solutions describe time-dependent masses of the form $m \, e^{i \o t}$. We do not have a good field theory intuition for the physics of such a mass term.

Third, all of the solutions that cross the semicircle are singular at $r=0$ and hence should be discarded as unphysical. We can see the singularity easily from fig. \ref{ysolzerob}: these solutions all have nonzero $y'(r)$ at $r=0$. Additionally, given our numerical solutions we have computed the Ricci scalar associated with the induced metric, eq. (\ref{nonzerospininducedmetric}), and observed the divergence explicitly.

In fig. \ref{abovecurv} we illustrate the behavior of the Ricci scalar for solutions ending above the semicircle. As $r \rarrow \infty$, we see $R(r) \rarrow -14$ for all solutions, the expected value for $AdS_5 \times S^3$. Solutions that end far above the semicircle, with $y(0) \gg \o$, should approach the constant solution $y(r) = c_0$ of a non-spinning D7-brane, and hence at $r=0$ should have $R(0) \rarrow -8$. In fig. \ref{abovecurv} we see that is the case. As $y(0)$ decreases toward $\o$, however, we see that the curvature at $r=0$ decreases, and appears to diverge when the D7-brane ends precisely at the semicircle, $y(0) = \o$. Such behavior is in fact familiar \cite{Babington:2003vm,Kirsch:2004km,Ghoroku:2005tf,Apreda:2005yz,Mateos:2006nu,Albash:2006ew,Karch:2006bv,Mateos:2007vn}: probe D7-branes in AdS-Schwarzschild may end ``above'' the black hole horizon or may intersect the horizon. The ``critical solution'' that ends precisely at the horizon is singular. We are seeing the same behavior, with the black hole horizon replaced by the semicircle.

\FIGURE{
\includegraphics[width=0.48\textwidth]{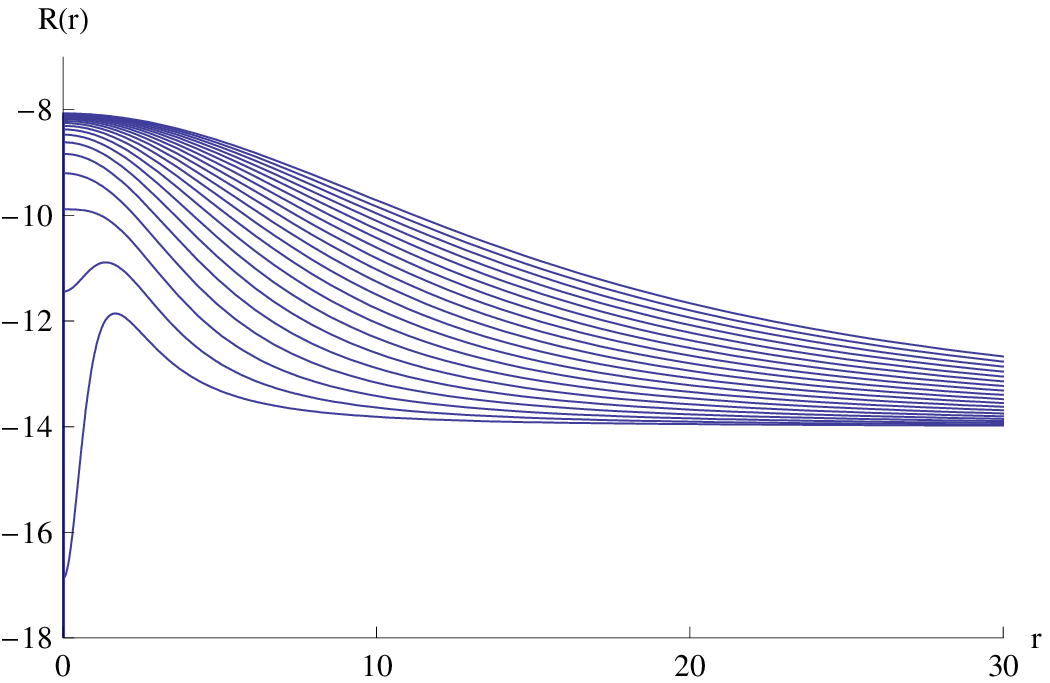} \hfil
\includegraphics[width=0.48\textwidth]{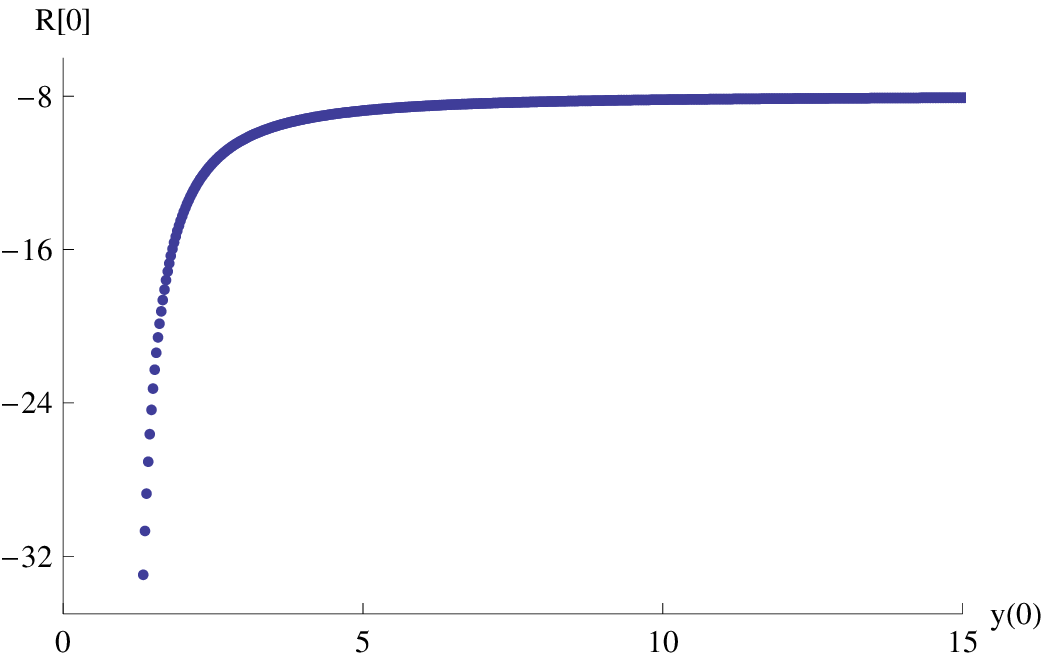}\\
(a) \hfil\hfil (b)
\caption{\label{abovecurv} The behavior of the Ricci scalar $R(r)$ for D7-branes that end above the semicircle of eq. (\ref{semicircle}), \textit{i.e.} for which $y(0) > \o$. We have used $\o=1$ to generate these figures. (a.) The Ricci scalar for solutions with various values of $y(0)$, ranging from $y(0) = \o + 15$ (the top curve) to $y(0) = \o + 0.76$ (the bottom curve). All solutions have $R(r) \rarrow -14$ as $r\rarrow \infty$. (b.) The Ricci scalar evaluated at $r=0$, $R(0)$, as a function of the position above the semicircle where the D7-brane ends, $y(0)$. We see that when $y(0) \gg \o$, $R(0) \approx -8$ as appropriate for the constant solution $y(r) = c_0$. As $y(0)$ approaches the top of the semicircle, $y(0) \rarrow \o$, we see that $R(0)$ diverges to negative infinity.}}

We will not present plots for the Ricci scalar of solutions that intersect the semicircle. We will only note that, again, as $r\rarrow \infty$, all solutions have $R(r) \rarrow -14$, and that the curvature of every solution diverges at $r=0$, as expected. Indeed, the curvature appears to be extremely large everywhere inside the semicircle. For example, setting $\o=1$, the solution intersecting the semicircle at the point $(r_0,y_0) \approx (0.29,0.91)$ reaches a curvature on the order of $10^{3}$ inside the semicircle within a distance $0.2$ of the semicircle, and a curvature of order $10^{7}$ at a distance of $0.4$. Clearly curvature corrections to the DBI action will be important for such solutions, so we cannot trust them. Nevertheless, we will include such solutions in our later analysis, for two reasons. First, these solutions are required to account for the full range of $c_0$ values. Second, we expect that, in an AdS-Schwarzschild background, the high-curvature region may be cloaked by the horizon, in which case such solutions may become physically acceptable. The analogous figures in refs. \cite{Erdmenger:2007bn,Albash:2007bq,Evans:2008zs} for D-branes in AdS-Schwarzschild suggest this.

As discussed in refs. \cite{Filev:2008xt,Evans:2008zs}, this system undergoes a first-order phase transition in which, roughly speaking, as $y(0)$ approaches $\o$ the D7-brane ``jumps'' from ending outside the semicircle to intersecting the semicircle. The transition is analogous to the D7-brane's first-order ``meson melting'' phase transition in the AdS-Schwarzschild background \cite{Babington:2003vm,Kirsch:2004km,Ghoroku:2005tf,Apreda:2005yz,Mateos:2006nu,Albash:2006ew,Karch:2006bv,Mateos:2007vn,Hoyos:2006gb} (for details, see refs. \cite{Filev:2008xt,Evans:2008zs}). The transition is between two solutions with nonzero $c_0$. We are interested only in solutions with $c_0=0$, so we will not investigate the phase transition here.

\section{Finite $U(1)_B$ Magnetic Field}
\label{u1b}

With $N_f$ massless flavor fields, the SYM theory has a global $U(N_f)$ symmetry. We identify the overall diagonal $U(1)$ as baryon number, $U(1)_B$. In the supergravity description, the $U(1)_B$ current is represented by the $U(1)$ gauge field, $A_{\mu}$, propagating on the D7-brane worldvolume. We can describe external electric and magnetic fields in the field theory, coupled to anything carrying $U(1)_B$ charge, by introducing non-normalizable modes for $A_{\mu}$ in the supergravity theory. For example, we will be interested in a magnetic field, which we introduce by adding to our D7-brane ansatz the constant field strength $F_{xy} = B$. In the SYM theory, we identify $F_{xy}$ as a constant $U(1)_B$ magnetic field pointing in the $z$ direction.

The utility of introducing $B$ is that, at zero temperature, zero mass, and zero $U(1)_R$ chemical potential, AdS/CFT calculations have shown that the $U(1)_B$ magnetic field triggers spontaneous breaking of the $U(1)_R$ symmetry \cite{Filev:2007gb,Filev:2007qu,Albash:2007bk,Erdmenger:2007bn}. In supergravity language, the D7-brane is ``repelled'' from the origin of the $(r,y)$ plane, so that the solution with zero asymptotic separation, $c_0 = 0$, is no longer just the trivial solution $y(r) = 0$.

To illustrate how this occurs, we will briefly review the results of refs. \cite{Filev:2007gb,Filev:2007qu,Albash:2007bk,Erdmenger:2007bn}. We consider an ansatz for the D7-brane fields with $y(r)$ and $F_{xy} = B$ only (so for now $\phi(t,r) = 0$). We will also define the notation $\tilde{B} \equiv (2\pi\alpha') B$. The D7-brane action is then
\beq
S_{D7} = \, - \N \, \int dr \, r^3 \, \sqrt{\left(1+y'(r)^2\right) \left( 1 + \frac{\tilde{B}^2}{(y^2 + r^2)^2} \right)}
\eeq
From the equation of motion, we find the asymptotic form of $y(r)$,
\beq
y(r) \, = \, c_0 \, + \, \frac{c_2}{r^2} \, + \, O\left(\frac{1}{r^4}\right),
\eeq
where again we translate to SYM theory quantities with $m = \frac{c_0}{2\pi\alpha'}$ and $\Omv \propto \, -2 \, c_2$.

We generate solutions numerically as follows. For all solutions we impose $y'(0) = 0$. We then choose the value of $y(0)$ and numerically integrate to large $r$. From these solutions we extract the values of $c_0$ and $c_2$.

We present the plot of $c_2$ as a function of $c_0$ in fig. \ref{nospinb} (a.). The curve actually spirals into the origin, crossing the vertical axis an infinite number of times. We thus have infinitely many solutions with $c_0=0$. As argued in ref. \cite{Erdmenger:2007bn}, however, the $c_0=0$ solution with lowest energy will be the physical one, which turns out to be the ``first'' $c_0=0$ solution, ``first'' meaning the first $c_0=0$ solution we reach as we enter the plot from the right (from large values of $c_0$). In fact, the other $c_0 = 0$ solutions are not only thermodynamically disfavored, they are unstable, having tachyonic fluctuations \cite{Filev:2007qu}. Notice in particular that the trivial solution $y(r)=0$, at the center of the spiral, is unstable.

\FIGURE{
\includegraphics[width=0.48\textwidth]{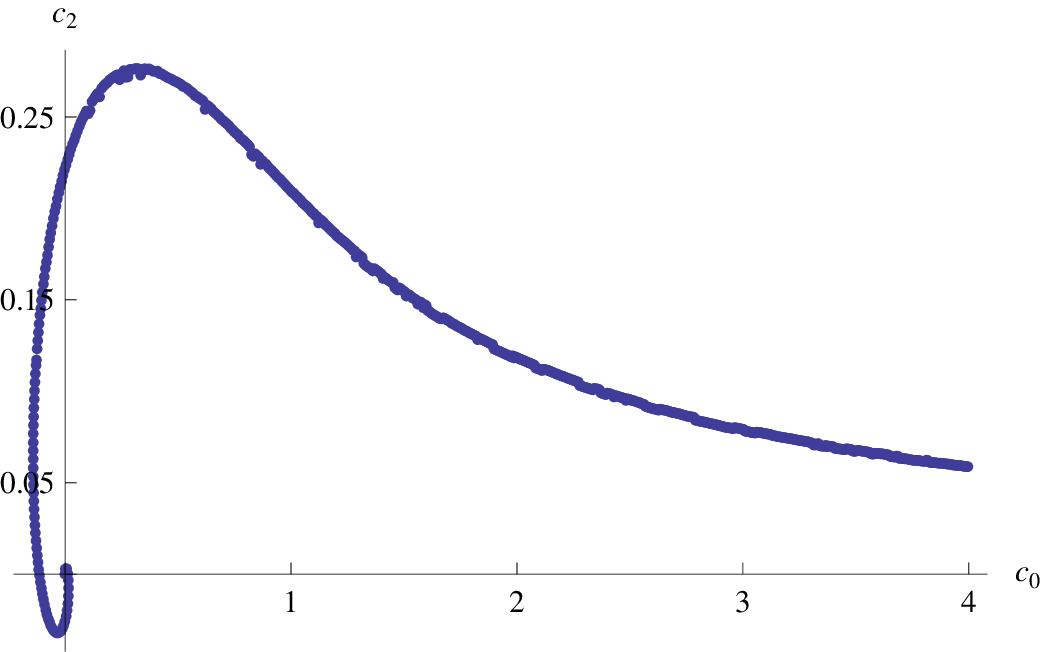} \hfil
\includegraphics[width=0.48\textwidth]{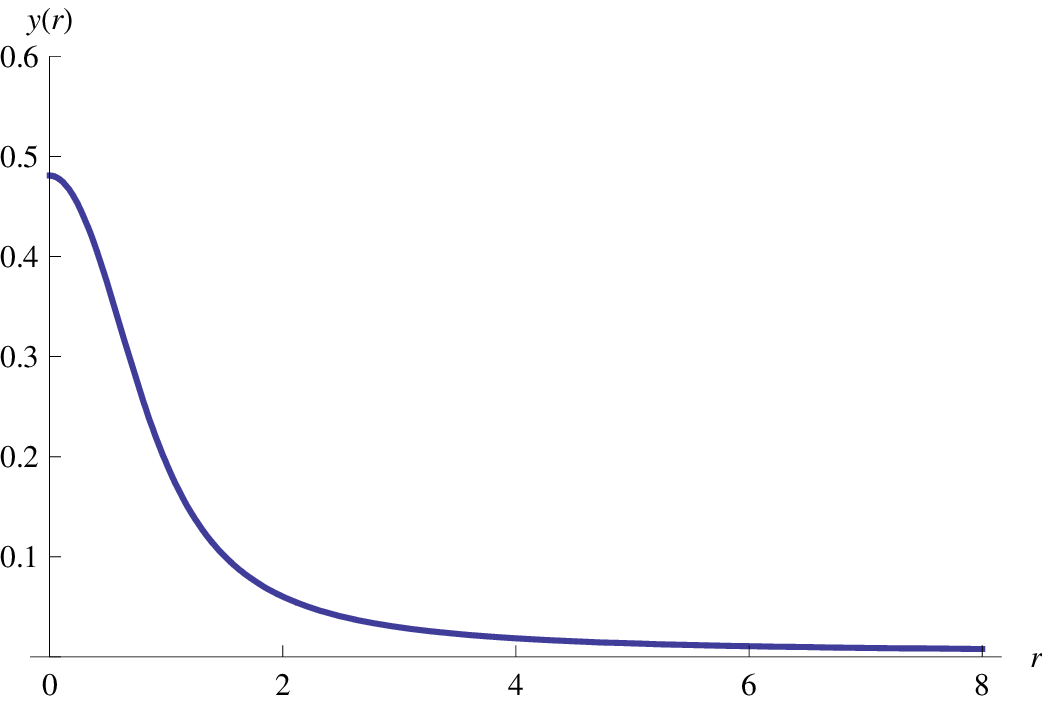}\\
(a) \hfil\hfil (b)
\caption{\label{nospinb} (a.) The value of the sub-leading asymptotic coefficient $c_2$ as a function of the leading coefficient $c_0$ for a D7-brane with nonzero worldvolume magnetic field $\tilde{B} = (2\pi\alpha')B$. We have used $\tilde{B} = 1$. The curve actually spirals into the origin, and hence crosses the $c_0=0$ (vertical) axis an infinite number of times, although that is not apparent in our plot. The physical $c_0=0$ solution has $c_2 = 0.226$. b.) The solution $y(r)$ for the physical $c_0=0$ solution, for $\tilde{B}=1$.}}

The physical $c_0=0$ solution has nonzero $c_2$, indicating that in the SYM theory the $U(1)_R$ is spontaneously broken. An analysis of the D7-brane's fluctuation spectrum, dual to the SYM theory's meson spectrum, confirmed the existence of a Goldstone boson associated with the symmetry breaking \cite{Filev:2007gb}. In the SYM theory, when $m=0$ the only scale in the problem is $B$, hence by dimensional analysis we have $\Omv \propto B^{3/2}$. More precisely, $\Omv = - \frac{1}{(2\pi)^3}\, \sqrt{\lambda} \nf \nc \, (2 \times 0.226) \, B^{3/2}$ \cite{Filev:2007gb}.

A picture of the physical $c_0=0$ solution in the $(r,y)$ plane appears in fig. \ref{nospinb} (b.). This is just what we want: a solution with zero asymptotic separation but nonzero extension into the $(r,y)$ plane. Our goal, roughly speaking, is to set this solution spinning, giving us a solution with zero asymptotic separation but nonzero angular momentum.

\section{Finite $U(1)_R$ Chemical Potential and $U(1)_B$ Magnetic Field}
\label{u1randu1b}

We will now study D7-branes spinning with angular frequency $\o$, and with a constant worldvolume magnetic field $F_{xy} = B$, which will produce embeddings with $c_0=0$ but $c_2 \neq 0$. Such solutions describe states in the SYM theory with massless hypermultiplet fields, a finite $U(1)_R$ density, and spontaneous breaking of the $U(1)_R$ symmetry.

We consider an ansatz for the D7-brane worldvolume fields with $y(r)$, $\phi(t,r) = \o t + f(r)$ and now $F_{xy} = B$. The DBI action becomes
\beq
S_{D7} \, = \, - \N \, \int dr \, r^3 \, \sqrt{\left( \left( 1 + y'^2 \right) \left( 1 - \dot{\phi}^2 \frac{y^2}{(y^2 + r^2)^2} \right) + y^2 \phi'^2 \right) \left( 1 + \frac{\tilde{B}^2}{(y^2 + r^2)^2} \right)}
\eeq
The constant of motion $c$ is now
\beq
\frac{\delta \lag}{\delta \phi'(r)} = \N r^3 \frac{y^2 \phi' \left( 1 + \frac{\tilde{B}^2}{(y^2 + r^2)^2} \right)}{\sqrt{\left( \left( 1 + y'^2 \right) \left( 1 - \o^2 \frac{y^2}{(y^2 + r^2)^2} \right) + y^2 \phi'^2 \right) \left( 1 + \frac{\tilde{B}^2}{(y^2 + r^2)^2} \right)}} \equiv c
\eeq
The solution for $\phi'(r)$ is now
\beq
\label{phisolnonzerob}
\phi'(r) = \frac{c}{y} \,  \sqrt{ \frac{\left( 1 + y'^2 \right) \left( 1 - \o^2 \frac{y^2}{(y^2 + r^2)^2} \right)}{\N^2 y^2 r^6 \left( 1 + \frac{\tilde{B}^2}{(y^2 + r^2)^2} \right) - c^2 } }
\eeq
Plugging this into the action gives
\beq
S_{D7} \, = \, - \N \, \int dr \, r^3 \, \left( 1 + \frac{\tilde{B}^2}{(y^2 + r^2)^2} \right) \, \sqrt{1+y'^2} \sqrt{\frac{1 - \o^2 \frac{y^2}{(y^2 + r^2)^2}}{1 + \frac{\tilde{B}^2}{(y^2 + r^2)^2} - \frac{c^2}{\N^2}\frac{1}{y^2 \, r^6}}}
\eeq
The Legendre transform of the action is
\bea
\hat{S}_{D7} & = & S_{D7} - \int dr \, \phi'(r) \, \frac{\delta S_{D7}}{\delta \phi'(r)} \nonumber \\ & = & -\N \int dr \, r^3 \sqrt{1+y'^2} \sqrt{\left( 1 - \o^2 \frac{y^2}{(y^2 + r^2)^2} \right) \left( 1 + \frac{\tilde{B}^2}{(y^2 + r^2)^2} - \frac{c^2}{\N^2} \frac{1}{y^2 r^6} \right)}.
\eea

We can see that the equation for the semicircle is unchanged, but the cubic curve has become
\beq
y(r) \, = \, \sqrt{\frac{c^2}{\N^2} \, \frac{1}{r^6} - \frac{\tilde{B}^2}{\o^2}}.
\eeq
The value of $c$ is still fixed uniquely by a point on the semicircle. The plot of $c$ versus $y$ is qualitatively similar to fig. \ref{c}.

The asymptotic forms of $y(r)$ and $\phi(t,r)$ are unchanged from the $\tilde{B}=0$ case, eqs. (\ref{spinningyasymptotic}) and (\ref{spinningphiasymptotic}). $\Omv$ and $\Ophiv$ are again given by eq. (\ref{spinningd7omv}) and $\Ophiv = c$, respectively.

We generate solutions numerically in precisely the same way as in section \ref{u1r}. We first consider solutions that intersect the semicircle. The behavior with nonzero $\tilde{B}$ is more complicated than with zero $\tilde{B}$, so in fig. \ref{circlesolseq} we present only a few examples. We choose two points on the semicircle and generate solutions with increasing $\tilde{B}$. As $\tilde{B}$ increases, for the solution intersecting the semicircle near the top, the value of $c_0$ first decreases, but then begins to increase. For the solution intersecting the semicircle near the bottom, the value of $c_0$ decreases, but only very little. We summarize the behavior of solutions with the three-dimensional plot in fig. \ref{3dmass}, where we plot $c_0$ as a function of $\tilde{B}$ and the value of $y$ where the solution intersects the semicircle, which we denote $y_0$.

\FIGURE{
\includegraphics[width=0.50\textwidth]{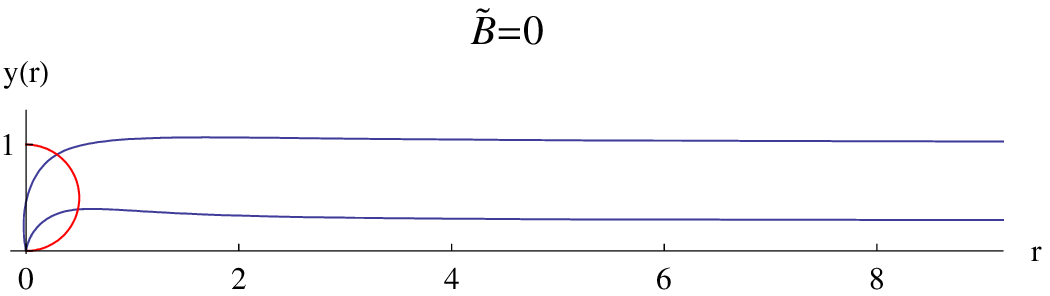}
\includegraphics[width=0.50\textwidth]{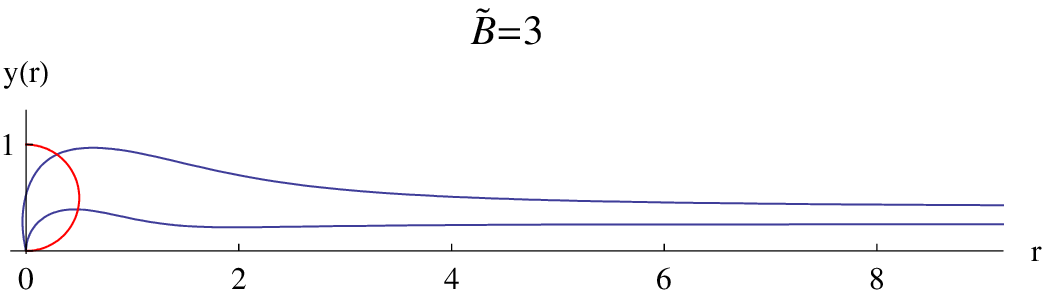}
\includegraphics[width=0.50\textwidth]{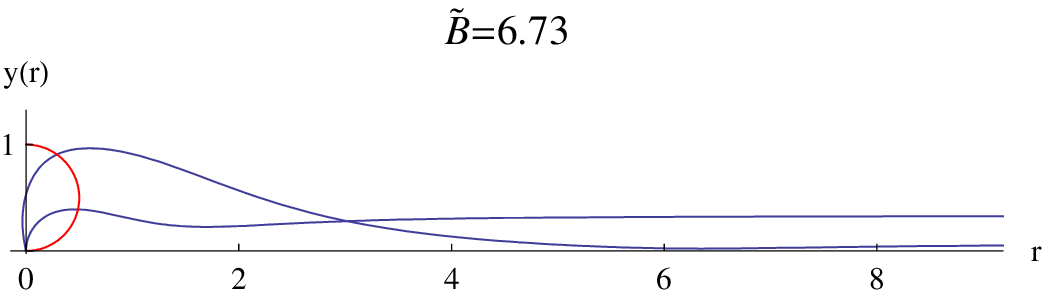}
\includegraphics[width=0.50\textwidth]{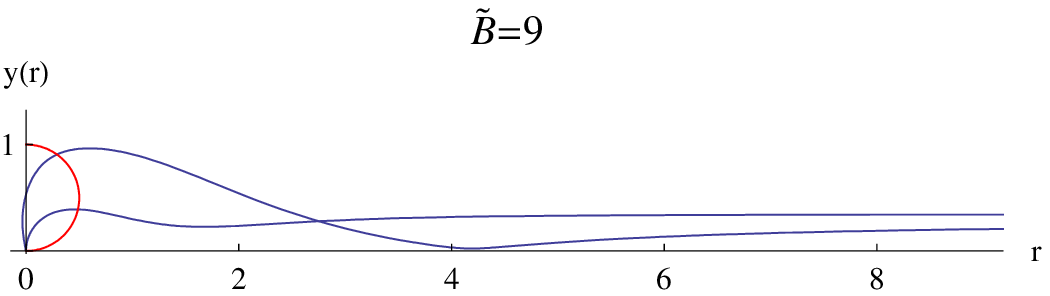}
\includegraphics[width=0.50\textwidth]{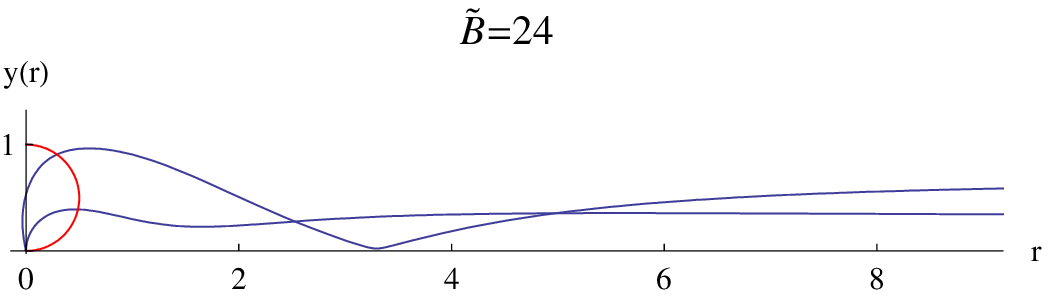}
\caption{\label{circlesolseq} Illustration of behavior of semicircle-intersecting solutions as $\tilde{B}$ increases. Here we set $\o=1$. The top figure has $\tilde{B} = 0$, and the subsequent figures show the behavior as $\tilde{B}$ increases. We present plots for the values $\tilde{B} = 3, 6.73$ (the critical value), $9$ and $24$.}
}

\FIGURE{
\includegraphics[width=0.70\textwidth]{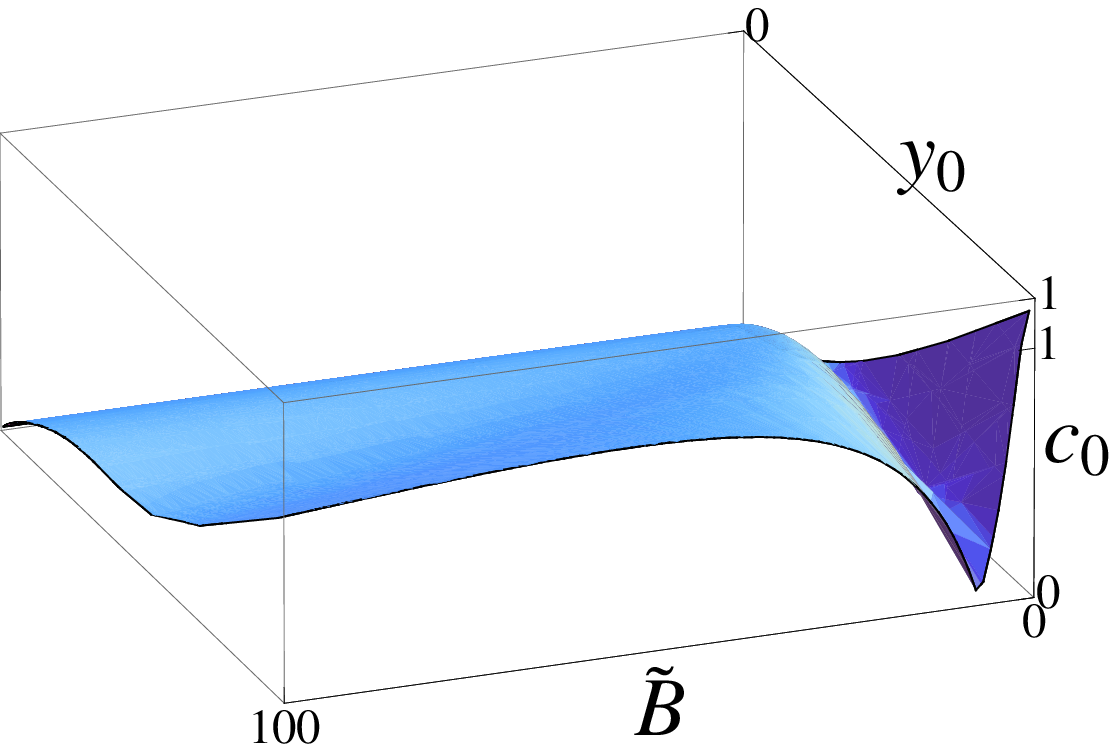}
\caption{\label{3dmass} Three-dimensional plot of $c_0$ as a function of $\tilde{B}$ and the value of $y$ where the solution intersects the semicircle, denoted $y_0$. Here we use $\o=1$. The only points of the surface that reach $c_0=0$ are at $y_0=0$, corresponding to the trivial solution (which intersects the semicircle at $y_0=0$).}}

The behavior of the Ricci scalar for the semicircle-intersecting solutions is qualitatively similar to the $\tilde{B} = 0$ case: for all the solutions that intersect the semicircle, $R(r)$ diverges at $r=0$, and indeed is extremely large almost everywhere inside the semicircle, and hence the solutions must be discarded as unphysical.

The principal result of figs. \ref{circlesolseq} and \ref{3dmass} is that all solutions that intersect the semicircle, which have nonzero $c$, also have nonzero $c_0$. We are only interested in solutions with $c_0=0$, however, so we will pay no more attention to solutions that intersect the semicircle.

Now for the solutions that end above the semicircle. We present some examples of such solutions in fig. \ref{abovesolseq}, for increasing values of $\tilde{B}$. Here we find that for a given point above the semicircle the corresponding value of $c_0$ \textit{decreases} as $\tilde{B}$ increases. Heuristically, as we increase $\tilde{B}$, the solutions ``bend down.'' Turning things around, if we imagine fixing $c_0$ and integrating into the bulk, then as we increase $\tilde{B}$ we see that the point (above the semicircle) where the solution reaches $r=0$ increases ($y(0)$ increases).

\FIGURE{
\begin{tabular}{cc}
\includegraphics[width=0.45\textwidth]{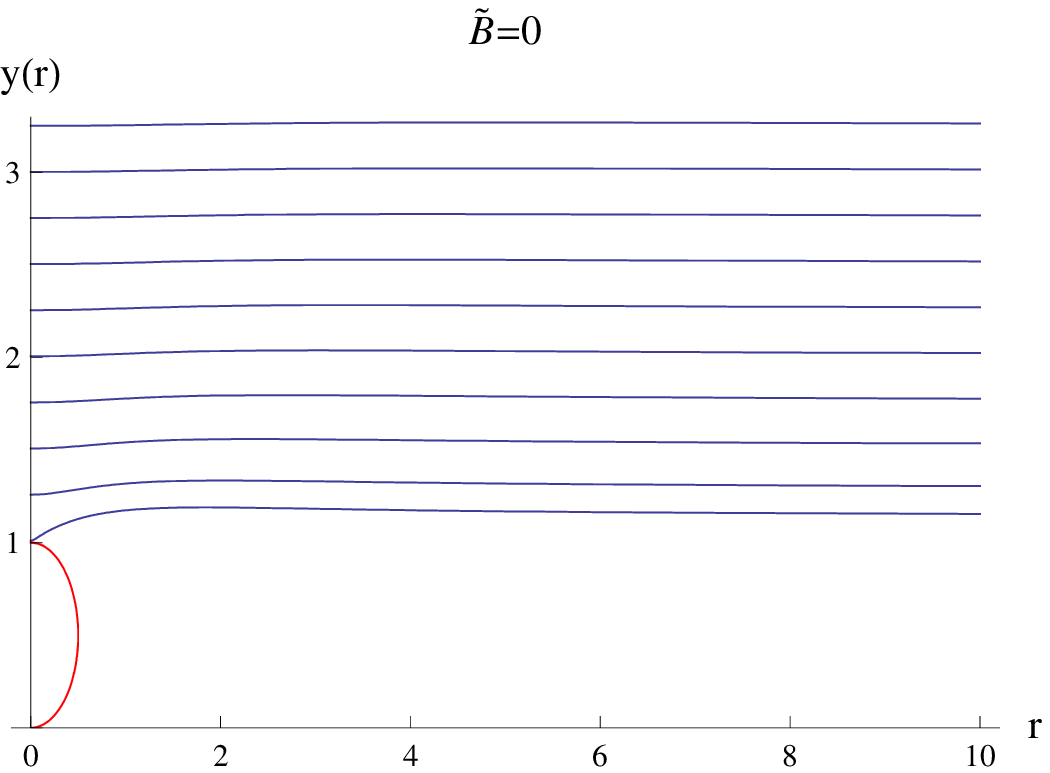} & \includegraphics[width=0.45\textwidth]{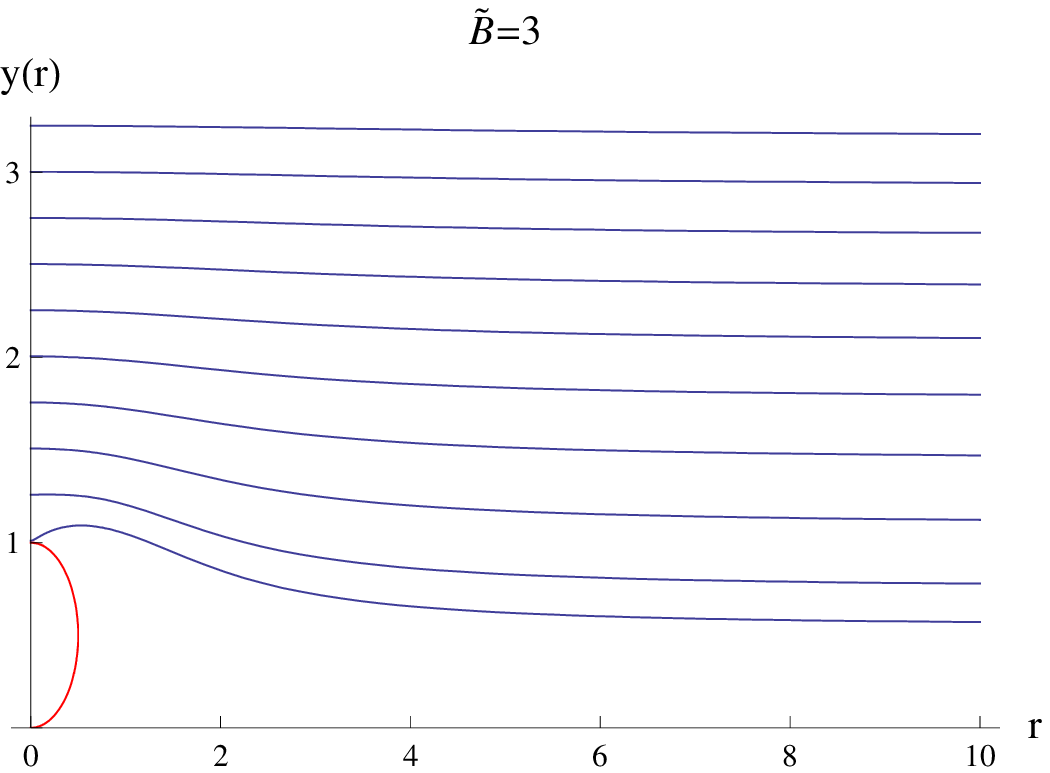}\\
\includegraphics[width=0.45\textwidth]{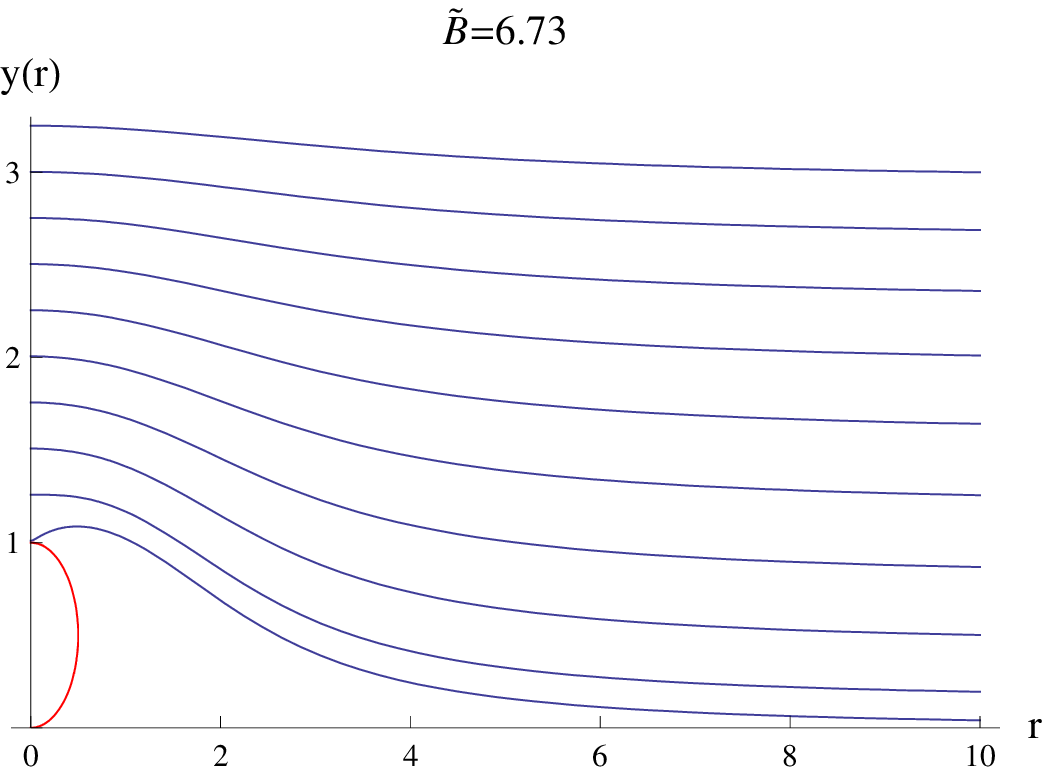} & \includegraphics[width=0.45\textwidth]{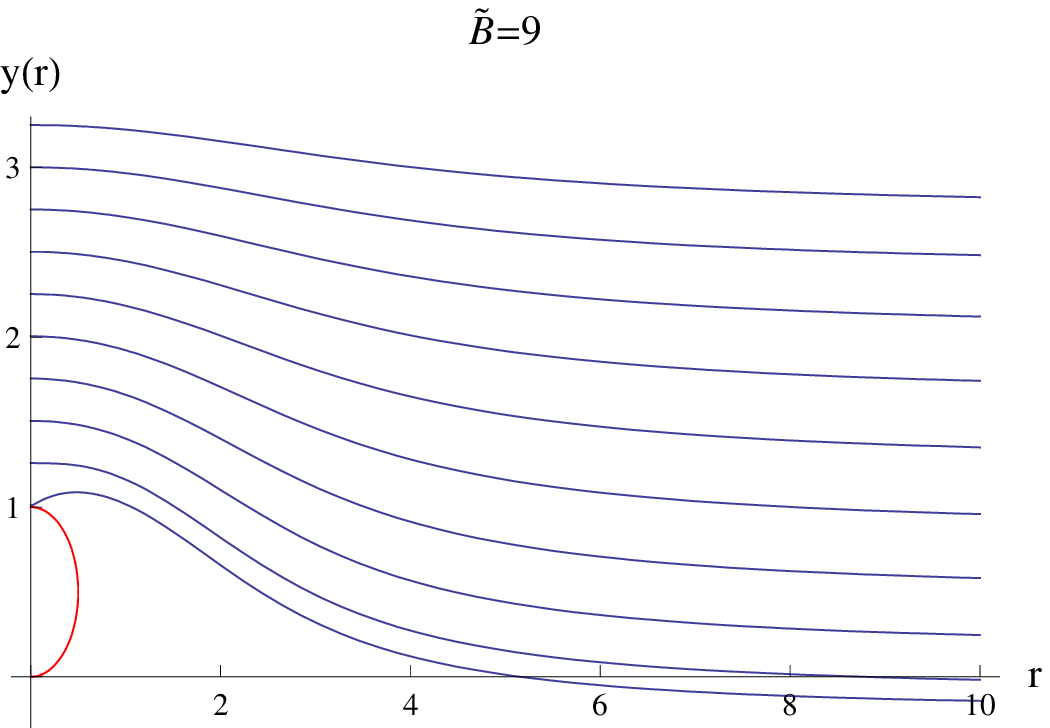}\\
\end{tabular}
\caption{\label{abovesolseq} Illustration of the behavior of solutions that reach $r=0$ above the semicircle. Here we have set $\o=1$. For a given value of $y(0)$, as we increase $\tilde{B}$ the associated value of $c_0$ decreases. We present plots for the values $\tilde{B} = 0, 3, 6.73$ and $9$. At the critical magnetic field $\tilde{B} \approx 6.73$ the first solution with $c_0=0$ appears.}
}

At a critical value of $\tilde{B} \approx 6.73$, a $c_0=0$ solution appears: see fig. \ref{abovesolseq}. A $c_0=0$ solution continues to exist for larger values of $\tilde{B}$, as we show in fig. \ref{zerosols} (a.). Indeed, if we look only at the $c_0=0$ solution, and increase $\tilde{B}$, we find that the solution's value of $y(0)$ increases. Heuristically, the D7-brane ``bends out'' more into the $y$ direction as $\tilde{B}$ grows. Notice that, from the SYM theory point of view, when $c_0=0$ the only scales in the problem are $\tilde{B}$ and $\o$, so when working with $c_0=0$ solutions we will always write $\tilde{B}$ in units of $\o$. The critical value of $\tilde{B}$ is thus $\tilde{B} \approx 6.73 \, \o^2$.

\FIGURE{
\begin{tabular}{cc}
\includegraphics[width=0.45\textwidth]{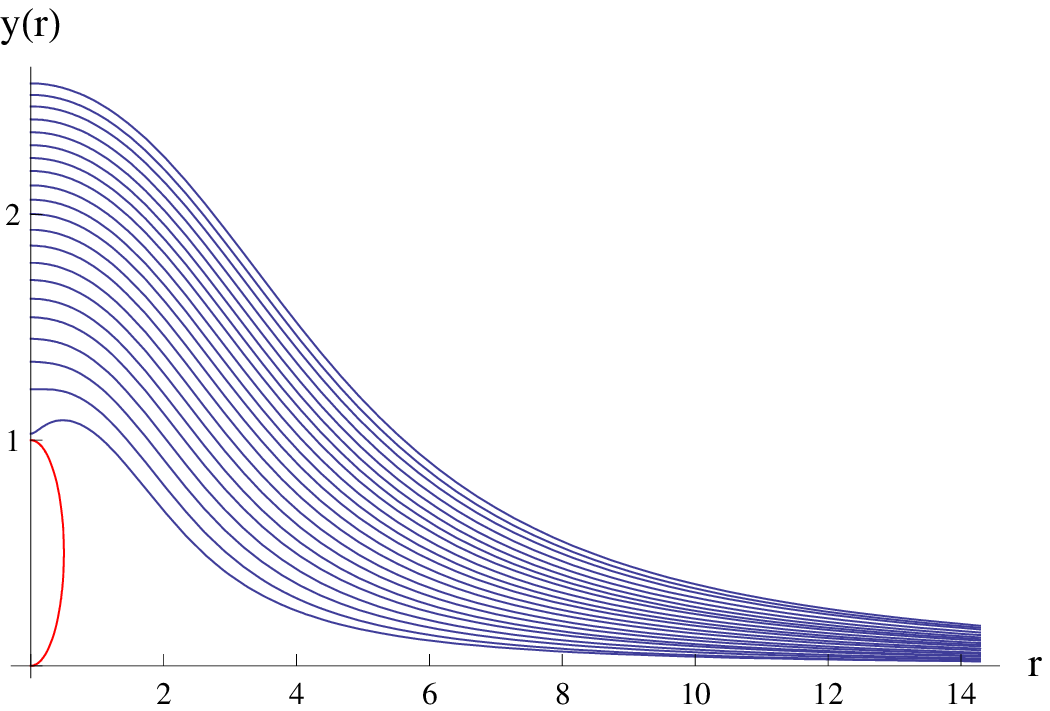} & \includegraphics[width=0.45\textwidth]{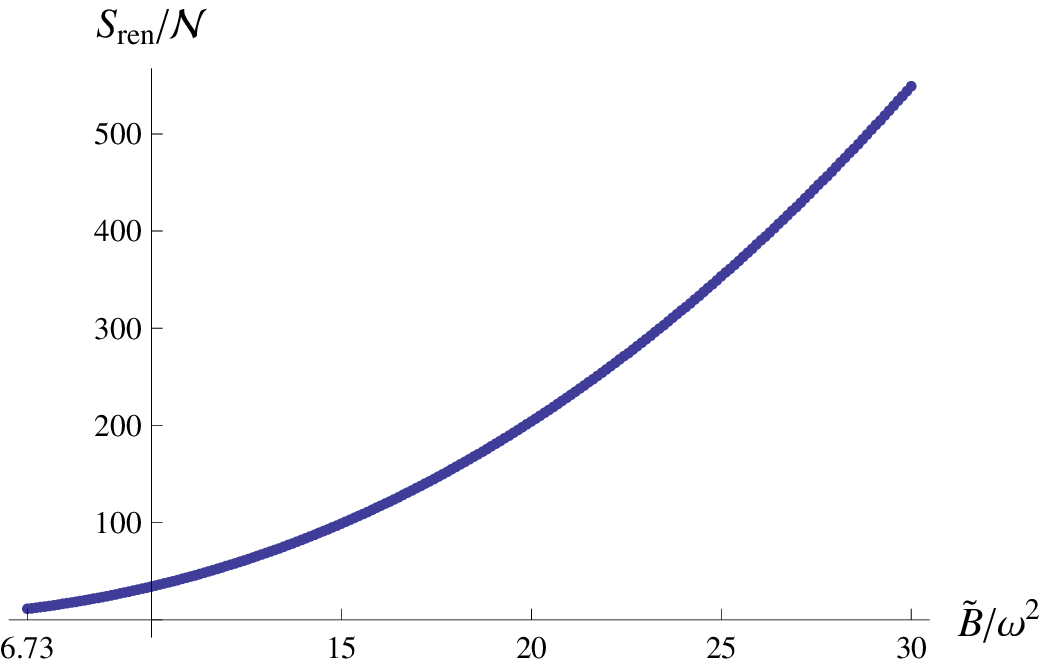}\\
(a.) & (b.)\\
\includegraphics[width=0.45\textwidth]{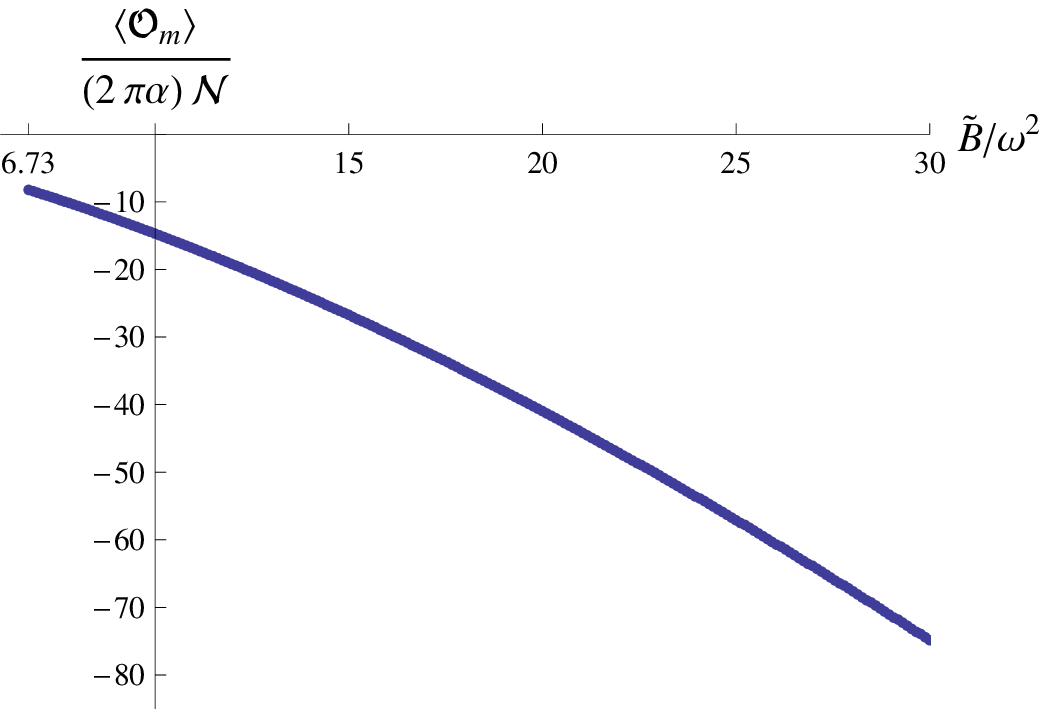} & \includegraphics[width=0.45\textwidth]{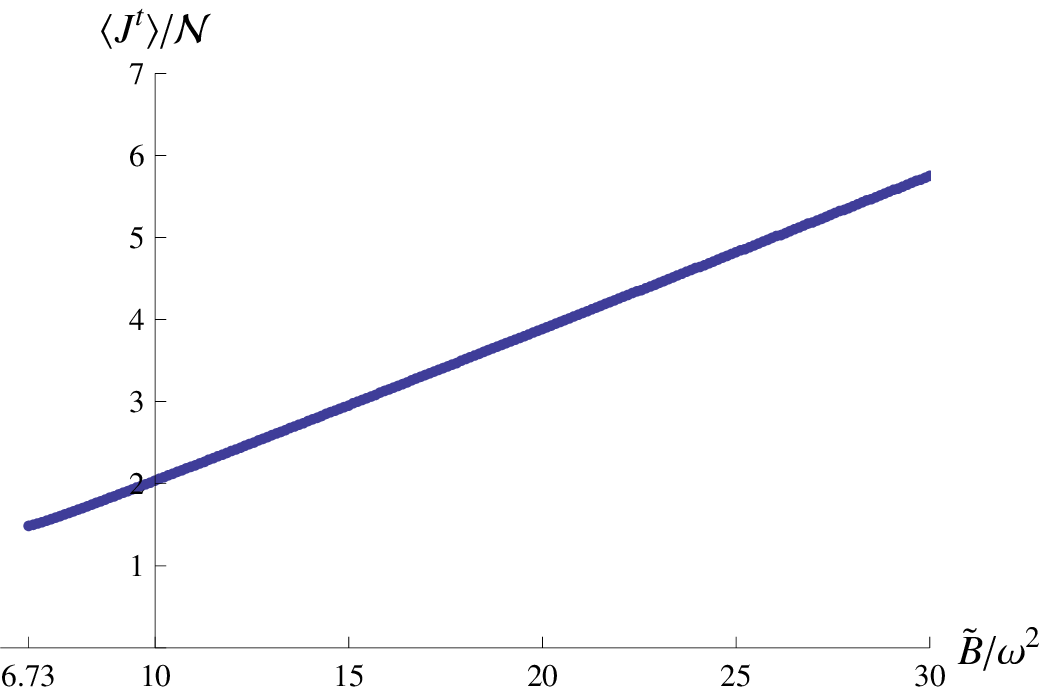}\\
(c.) & (d.)\\
\end{tabular}
\caption{\label{zerosols} (a.) The behavior of solutions with $c_0=0$ as $\tilde{B}/\o^2$ increases. We use values of $\tilde{B}/\o^2$ ranging from $\tilde{B}/\o^2 = 6.73$ (bottom curve) to $\tilde{B}/\o^2 = 30$ (top curve). As $\tilde{B}/\o^2$ grows, the associated value of $y(0)$ increases: the D7-brane ``bends out'' more in the $y$ direction. (b.) The value of the renormalized D7-brane action divided by $\N$ as a function of $\tilde{B}/\o^2$ for solutions with $c_0=0$. The value of $S_{ren}/\N$ at $\tilde{B}/\o^2=6.73$ is approximately $11.3$. (c.) The expectation value $\Omv$, divided by $(2\pi\alpha')\N$, as a function of $\tilde{B}/\o^2$ for solutions with $c_0=0$. (d.) The density $\jtv$ divided by $\N$ as a function of $\tilde{B}/\o^2$ for solutions with $c_0=0$.}
}

The trivial solution $y(r)=0$ is another $c_0=0$ solution, so once $\tilde{B}/\o^2$ reaches the critical value we can make a meaningful comparison between two $c_0=0$ solutions. To determine which is preferred, we must, in supergravity language, compare the values of their on-shell action $S_{D7}$, or, in SYM theory language, compare the values of their free energy $\O$. Recall that we identify $S_{D7} = - \O$, so that the solution with \textit{larger} $S_{D7}$ will be thermodynamically preferred.

The on-shell action $S_{D7}$ suffers from divergences coming from the integration over the infinite volume of $AdS_5$. In SYM theory language, these are UV divergences, which we can cancel with counterterms. In the Appendix we perform the ``holographic renormalization" of the on-shell action by regulating and then cancelling the divergences using counterterms. We will denote the renormalized action $S_{ren}$, and identify $S_{ren} = - \O$.

The trivial solution has $S_{ren}=0$. We find numerically that when $\tilde{B}/\o^2 = 6.73$, the non-trivial $c_0=0$ solution has $S_{ren}/\N \approx 11.3 > 0$, and hence the nontrivial solution is thermodynamically preferred. The value of $S_{ren}$ for the nontrivial $c_0=0$ solution increases monotonically with $\tilde{B}/\o^2$ as shown in fig. \ref{zerosols} (b.), so the non-trivial solution remains the preferred solution as we increase $\tilde{B}/\o^2$.

Na\"{i}vely, then, we might think the system exhibits a first-order phase transition. In supergravity language, the D7-brane ``jumps'' from the trivial embedding to a nontrivial embedding, with nonzero angular momentum. In SYM theory language, the theory jumps from a state in which $\Omv = \jtv = 0$ to a state with nonzero $\Omv$ and $\jtv$. In fig. \ref{zerosols} (c.) we plot $\Omv$ (divided by $(2\pi\alpha')\N$) as a function of $\tilde{B}/\o^2$ and in fig. \ref{zerosols} (d.) we plot $\jtv$ divided by $\N$ as a function of $\tilde{B}/\o^2$, both for solutions with $c_0=0$. Clearly both are nonzero at the critical value $\tilde{B}/\o^2 = 6.73$. As $\Omv$ and $\jtv$ are first derivatives of $\O$, we seem to have a first-order transition.

That is not correct, however, because the free energy \textit{itself} is discontinuous: it jumps from zero to nonzero at the critical value of $\tilde{B}/\o^2$. Such behavior is unphysical, and signals to us that something is missing. More specifically, some class of $c_0=0$ embeddings appears to be absent for values of $\tilde{B}/\o^2$ below the critical value. What kind of embeddings could ``fill the gap'' is not obvious to us, so we leave this as an open question.

Moreover, we should not trust the nontrivial embedding precisely at the critical value of $\tilde{B}/\o^2$ because such an embedding has very high curvature. Indeed, the story of the scalar curvature for embeddings that end above the circle is qualitatively the same as in the $\tilde{B}=0$ case: the curvature is finite everywhere, but the curvature at $r=0$ grows as $y(0)$ approaches $\o$ (the top of the circle), where it diverges. In fig. \ref{curvzerom} (a.) we plot the Ricci scalar for the $c_0=0$ solutions as we increase the value of $\tilde{B}/\o^2$. For the critical value $\tilde{B}/\o^2$, the curvature of $r=0$ diverges to negative infinity. In fig. \ref{curvzerom} (b.) we plot the value of the Ricci scalar at the $r=0$ endpoint for the $c_0=0$ solutions as a function of $\tilde{B}/\o^2$. We see again that the curvature diverges at $r=0$ for the critical solution and then increases as we increase $\tilde{B}/\o^2$. The closer we come to the critical solution, the less we should trust our solutions.

\FIGURE{
\includegraphics[width=0.48\textwidth]{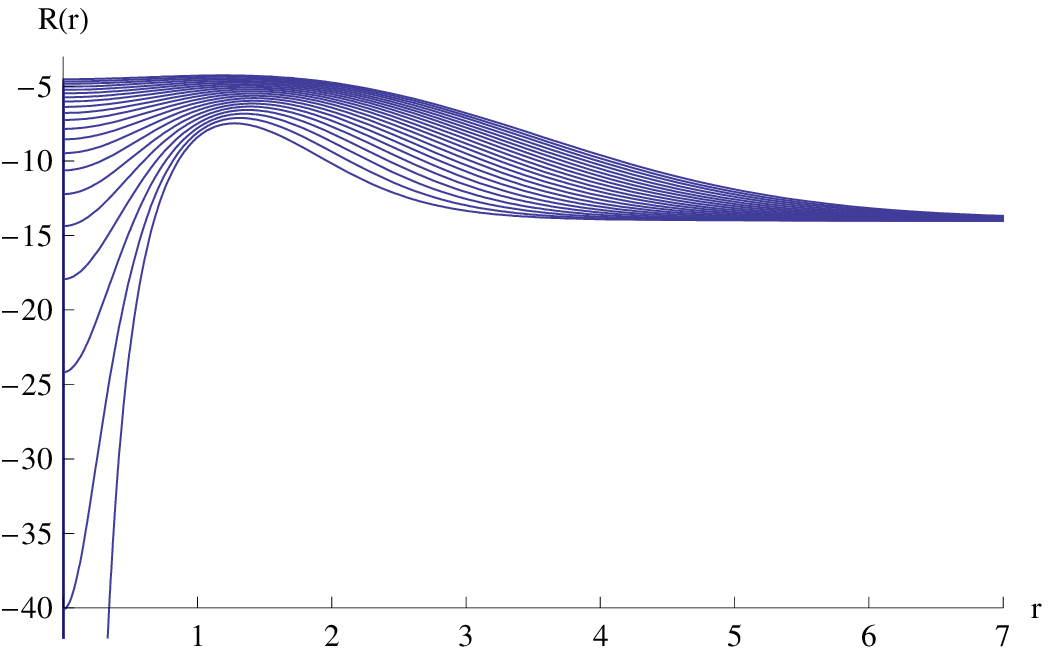} \hfil
\includegraphics[width=0.48\textwidth]{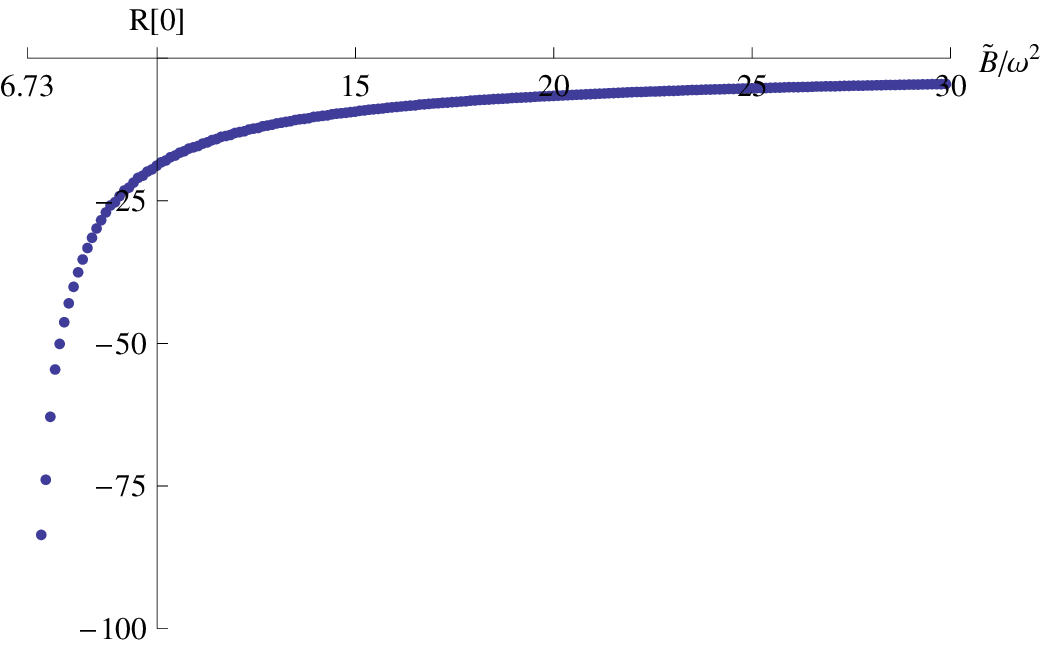}\\
(a) \hfil\hfil (b)
\caption{\label{curvzerom} The behavior of the Ricci scalar for D7-brane solutions with finite $\o$ and $\tilde{B}$ and with $c_0=0$. (a.) The Ricci scalar $R(r)$ for the $c_0=0$ solutions for values of $\tilde{B}/\o^2$ ranging from $\tilde{B}/\o^2 = 6.73$ (the bottom curve) up to $\tilde{B}/\o^2 = 30$ (the top curve). (b.) The value of the Ricci scalar at the $r=0$ endpoint, $R(0)$, for $c_0=0$ solutions as a function of $\tilde{B}/\o^2$.}}

\section{Conclusion}
\label{conclusion}

We have numerically constructed solutions, reliable within the supergravity approximation, for a spinning D7-brane with a worldvolume magnetic field embedded in $AdS_5 \times S^5$. These solutions describe $\N=4$ SYM theory coupled to massless $\N=2$ hypermultiplets in a state with a nonzero background $U(1)_B$ magnetic field and nonzero $U(1)_R$ charge. The $U(1)_B$ magnetic field causes spontaneous breaking of the $U(1)_R$ symmetry, hence the system should exhibit $U(1)_R$ superconductivity. We initiated the study of the zero-temperature thermodynamics of the system, and determined that for large enough values of the magnetic field the system prefers a state of broken symmetry. For smaller values we found that our class of D7-brane embeddings was insufficient to describe all equilibrium states of the SYM theory. We will end with some suggestions for future research\footnote{In our suggestions we continue to ignore the chemical potential of the adjoint fields. Of course a worthwhile extension would be to include the adjoint fields' chemical potential, and to study the system on a spatial three-sphere, which stabilizes the theory for sufficiently small chemical potential \cite{Yamada:2006rx,Yamada:2008em,Hollowood:2008gp}. A good question, in supergravity language, is whether our spinning D7-brane with worldvolume magnetic field minimizes the on-shell action, as opposed to a state in which the background geometry carries the angular momentum.}.

The biggest open question is, of course, where are the ``missing'' $c_0=0$ embeddings for small values of $\tilde{B}/\o^2$, which must fill the ``gap'' in the free energy that we discovered in section \ref{u1randu1b}? We have left this as an open problem, being content that we found solutions with the properties we wanted ($c_0=0$, nonzero angular momentum, and finite curvature).

An important generalization would be to introduce a finite temperature $T$, corresponding to D7-branes spinning in an AdS-Schwarzschild black hole background \cite{Witten:1998zw}. We have hope that many of the singular embeddings we found would be ``cured,'' in the sense that the high-curvature region would fall behind the black hole horizon. Additionally, in refs. \cite{Albash:2007bk,Erdmenger:2007bn}, the finite-temperature physics of D7-branes with zero $\o$ but nonzero $B$ was studied, with the result in the SYM theory that at high temperature the $U(1)_R$ symmetry is restored. We expect something similar to happen when we include nonzero $\o$. In SYM theory language, such a transition should extinguish the $U(1)_R$ superconductivity.

Similarly, another important generalization would be to introduce a background magnetic field for the $U(1)_R$ symmetry. We expect on general grounds that a sufficiently large magnetic field should extinguish superconductivity, that is, should restore the $U(1)_R$ symmetry. More generally, the phase diagram in the plane of $U(1)_R$ magnetic field versus temperature should be explored.

The fluctuation spectrum of spinning D7-branes should be computed, which should reveal the effects of a finite $U(1)_R$ chemical potential on the meson spectrum of the SYM theory. Such an analysis should also exhibit explicitly the Goldstone boson associated with the breaking of the $U(1)_R$ symmetry.

We chose an ansatz for the worldvolume fields that preserved many symmetries, such as translation invariance in the field theory directions. An interesting extension would be to consider a more general ansatz, respecting fewer symmetries. Indeed, in QCD at low temperature, asymptotically high baryon number chemical potential, and in the large-$\nc$ limit (with $\nf$ fixed), the ground state may break translation invariance, forming a so-called ``chiral density wave'' \cite{Deryagin:1992rw,Shuster:1999tn}.

We have claimed that our system describes a superconductor, so perhaps the most exciting task for the future would be studying the transport properties associated with the $U(1)_R$ charge and exhibiting superconductivity explicitly. We expect, for example, to see a gap in the frequency dependence of the $U(1)_R$ conductivity.

\section*{Acknowledgments}

We would like to thank F. Rust, J. Shock, D.T. Son, and L. Yaffe for many useful discussions, and M. Ammon, J. Erdmenger, A. Karch, M. Kaminski, R. Meyer, and D. Yamada for reading and commenting on an earlier version of the manuscript. We especially want to thank the Perimeter Institute for hospitality, and R. Myers for collaboration, during the early stages of this project. This work was supported in part by the Cluster of Excellence ``Origin and Structure of the Universe.''

\begin{appendix}

\section*{Appendix: Holographic Dictionary}
\label{holodic}

In this appendix we will: 1.) write explicitly the SYM theory operators dual to the D7-brane worldvolume fields $y$ and $\phi$, 2.) regulate and renormalize the on-shell D7-brane action, and 3.) compute the expectation values of the SYM theory operators dual to $y$ and $\phi$, as well as the expectation value $\jtv$, from the renormalized D7-brane action.

First, we will identify the operators dual to $y$ and $\phi$. These have been identified in several references; we borrow the results of ref. \cite{Myers:2007we}. We decompose the $\N=2$ hypermultiplet into two $\N=1$ chiral multiplets of opposite chirality. Let $\psi$ and $q$ denote the Weyl fermion and complex scalar of one chiral multiplet, and $\tilde{\psi}$ and $\tilde{q}$ the Weyl fermion and complex scalar of the other chiral mutliplet. In particular, $\psi$ and $\tilde{\psi}$ have opposite chirality.

The operator dual to $y$ is the supersymmetric completion of the mass operator, which we denote $\Om$. In terms of the SYM theory fields, $\Om$ is
\beq
\Om = i \tilde{\psi} \psi + \tilde{q} \left( m + \sqrt{2} \, \Phi \right) \tilde{q}^{\dagger} + q \left( m + \sqrt{2} \, \Phi \right) q^{\dagger} \, + \, h.c.
\eeq
Here we use the notation $\Phi$ to denote the complex scalar of the $\N=4$ multiplet with $U(1)_R$ charge $+2$.

The field $\phi$ is dual to the phase of $\Om$ (\textit{i.e.} fluctuations of $\phi$ are dual to fluctuations of the phase of the hypermultiplet mass term). We denote this operator as $\Ophi$. In terms of SYM theory fields, $\Ophi$ is
\beq
\Ophi = \tilde{\psi} \psi + i \sqrt{2} \, \tilde{q} \, \Phi \, \tilde{q}^{\dagger} + i \sqrt{2} \, q \, \Phi \, q^{\dagger} \, + \, h.c.
\eeq

Next, we will show how to compute finite on-shell D7-brane actions. The on-shell action diverges due to integration over the radial coordinate, $r$ all the way to the $AdS_5$ boundary at $r=\infty$. To obtain a finite on-shell action, we first regulate the integral by introducing a cutoff, $r = \L$. We then introduce counterterms localized at the $r = \L$ hypersurface to cancel the divergences of the action, and then send $\L \rarrow \infty$, yielding a finite result. This procedure is called ``holographic renormalization'' \cite{Henningson:1998gx,Henningson:1998ey,Balasubramanian:1999re,deHaro:2000xn,Skenderis:2002wp,Karch:2005ms}. We will denote the induced metric on the $r=\L$ hypersurface as $\g_{\mu \nu}$,
\beq
ds^2_{r=\L} \, = \, \g_{\mu \nu} \, dx^{\mu} dx^{\nu} \, = \, \L^2 \, \eta_{\mu \nu} dx^{\mu} dx^{\nu}
\eeq
and it determinant as simply $\g$, so that $\sqrt{-\g} = \L^4$.

We will give $\phi$ the general coordinate dependence
\beq
\phi(x,r) = k \cdot x + f(r)
\eeq
with $x_{\mu}$ the coordinates of (3+1)-dimensional Minkowski space (hence $\mu$ runs from $0$ to $3$) and $k_{\mu}$ a four-vector: $k_{\mu} = (-\o,\vec{k})$ with spatial vector $\vec{k}$. We will denote purely spatial indices with lower-case Latin indices, for example, a component of $\vec{k}$ will be $k_i$. We will also use the notation $\left( \partial \phi \right)^2 = \eta^{\mu \nu} \, \partial_{\mu} \phi \, \partial_{\nu} \phi$.

We denote the regulated D7-brane action as $S_{reg}$, so that $S_{reg} = - \int^{\L} dr \, \lag$. With the ansatz $y(r)$, $\phi(x,r)$ and $F_{xy} = B$, we have
\beq
\label{sreg1}
S_{reg} = - \N \, \int^{\L}dr \, r^3 \sqrt{\left( \left( 1+ y'^2\right) \left( 1 + \left( \partial \phi \right)^2 \frac{y^2}{(r^2 + y^2)^2} \right) + y^2 \phi'^2 \right) \left( 1 + \frac{\tilde{B}^2}{(y^2 + r^2)^2} \right)}
\eeq
Inserting the asymptotic forms of the solutions in eqs. (\ref{spinningyasymptotic}) and (\ref{spinningphiasymptotic}) (with $\o^2 \rarrow - \left( \partial \phi \right)^2$), we find
\bea
\label{sreg2}
S_{reg} & = & - \N \, \int^{\L}dr \, \left[  \, r^3 \, + \, \frac{1}{2} \, c_0^2 \, \left( \partial \phi \right)^2 \frac{1}{r} \, + \, \frac{1}{2} \, \tilde{B}^2 \frac{1}{r} \, + \, O\left( \frac{\log r}{r^2} \right) \right] \nonumber \\ & = & - \N \, \left [ \frac{1}{4} \, \Lambda^4 + \frac{1}{4} \, c_0^2 \, \left(\partial \phi \right)^2 \, \log \Lambda^2 + \frac{1}{4} \, \tilde{B}^2 \log \Lambda^2 + O\left( \frac{\log \Lambda}{\Lambda^2} \right) \right]
\eea
We then introduce counterterms on the $r = \L$ hypersurface to cancel the divergences. The counterterms that we use throughout this paper are
\begin{subequations}
\label{counterterms}
\beq
L_1 \, = \, + \frac{1}{4} \, \N \, \sqrt{-\g}
\eeq
\beq
L_2 \, = \, - \frac{1}{4} \, \N \sqrt{-\g} \, \left( \g^{\mu \nu} \partial_{\mu} \phi \, \partial_{\nu} \phi \right)\, g_{\phi \phi}(\L) \, \left( \log g_{\phi \phi}(\L) \, + \, 1 \right)
\eeq
\beq
L_3 \, = \, + \frac{1}{8} \, \N \, \sqrt{-\g} \, (2 \pi \alpha')^2 \, F^{ij} F_{ij} \, \log (\L^2)
\eeq
\end{subequations}
where $g_{\phi \phi}(\L) = \frac{y(\L)^2}{\L^2}$ and $\tilde{F}_{ij}$ is $(2\pi\alpha')$ times the D7-brane worldvolume field strength, which for us will include only $\tilde{F}_{xy} = \tilde{B}$. Written more explicitly, the counterterms are
\begin{subequations}
\label{explicitcounterterms}
\beq
L_1 \, = \, + \frac{1}{4} \, \N \, \frac{\L^4}{L^4} \nonumber
\eeq
\beq
L_2 \, = \, - \frac{1}{4} \, \N \, c_0^2 \, \left( \partial \phi \right)^2 \, \left( \log (c_0^2) \, - \, \log (\L^2) \, +1 \right) \, + \, O\left( \frac{\log \L}{\L^2} \right) \nonumber
\eeq
\beq
L_3 \, = \, + \frac{1}{4} \, \N \, \tilde{B}^2 \, \log (\L^2)
\eeq
\end{subequations}
where the divergences in $\L$ cancel those in the regulated action, eq. (\ref{sreg2}). The renormalized on-shell action, $S_{ren}$, is
\beq
S_{ren} = \lim_{\L \rarrow \infty} \left( S_{reg} + \sum_i L_i \right).
\eeq

A number of finite counterterms are possible. Indeed, some of the terms in eqs. (\ref{counterterms}) are finite. We have chosen the particular counterterms above so that $\Omv$ will have the appropriate behavior as $c_0 \rarrow \infty$: in SYM theory language, when $m \rarrow \infty$ the flavor fields decouple from the dynamics of the SYM theory. We must have $\Omv \rarrow 0$ in this limit, which fixes the finite counterterms. We will ignore all other possible finite counterterms (\textit{i.e.} we will set their coefficients to zero).

We will now compute the expectation values $\Omv$ and $\Ophiv$. In the AdS/CFT correspondence \cite{Maldacena:1997re,Gubser:1998bc,Witten:1998qj}, we identify the on-shell supergravity action with the generating functional (or grand canonical potential) of the SYM theory as $S_{ren} = - \Omega$. We thus have
\beq
\Omv \, = \, \frac{\delta \Omega}{\delta m} \, = \, - (2\pi\alpha') \frac{\delta S_{ren}}{\delta y(\L)} \, = \, -(2\pi\alpha') \, \lim_{\L \rarrow \infty} \left( \frac{\delta S_{reg}}{\delta y(\L)} + \sum_i \frac{\delta L_i}{\delta y(\L)}\right)
\eeq
The contribution from $S_{reg}$ is
\beq
\label{dldyprime}
\frac{\delta S_{reg}}{\delta y(\L)} \, = \, - \left . \frac{\delta \lag}{\delta y'} \right |_{r=\L} = \N \left ( 2 \, c_2 \, + \, \frac{1}{2} \left( \partial \phi \right)^2 \, c_0 \, \left( 1 -\log (\L^2) \right) \, + \, O\left( \frac{\log \L}{\L^2} \right) \right)
\eeq
where in the second equality we have inserted the asymptotic solutions eqs. (\ref{spinningyasymptotic}) and (\ref{spinningphiasymptotic}). Of the counterterms, $L_1$ and $L_3$ contribute nothing, while the contribution from $L_2$ is
\bea
\frac{\delta L_2}{\delta y(\L)} & = & - \frac{1}{2} \, \N \left( \partial \phi \right)^2 y(\L) \left[ \log\left( \frac{y(\L)^2}{\L^2}\right) + 2 \right] \\ & = & - \frac{1}{2} \,\N \, \left( \partial \phi \right)^2 \, c_0 \, \left[ \log(c_0^2) - \log (\L^2) + 2 \right] \, + \, O\left( \frac{\log \L}{\L^2} \right)
\eea
Summing everything and taking $\L \rarrow \infty$, we find
\beq
\label{omvformula}
\Omv \, = \, -(2 \pi \alpha') \, \N \, \left( \, 2 \, c_2 \, - \, \frac{1}{2} \, \left( \partial \phi \right)^2 \, c_0 \, - \frac{1}{2} \,  \left( \partial \phi \right)^2 \, c_0 \, \log (c_0^2)  \right)
\eeq
In terms of SYM theory quantities, the prefactor is $(2 \pi \alpha') \, \N = \frac{1}{(2\pi)^3} \, \sqrt{\lambda} \, \nf \, \nc$.

To show that $\Omv \rarrow 0$ as $c_0 \rarrow \infty$, we need to know the large-$c_0$ behavior of $c_2$. To determine this, we borrow arguments of ref. \cite{Filev:2007gb}: to study large $c_0$, we let $y(r) = c_0 + Y(r)$ and linearize the $Y(r)$ equation of motion, retaining only the leading terms in $(c_0^2 + r^2)^{-1}$. The equation of motion for the fluctuation $Y(r)$ is then
\beq
\partial_r \left[ r^2 \, Y'(r) \right] \, + \, r^3 \left[ \frac{2 c_0 \tilde{B}^2 + c_0 \left( \partial \phi \right)^2 (c_0^2 - r^2) }{(c_0^2 + r^2)^3} \right] \, = \, 0
\eeq
which has the solution
\beq
Y(r) = \alpha_1 + \frac{\alpha_2}{r^2} - \frac{1}{4} \frac{c_0 \left( \partial \phi \right)^2 (r^2 + 2c_0^2)}{r^2 (c_0^2 + r^2)} - \frac{1}{4} c_0 \left( \partial \phi \right)^2 \frac{\log(c_0^2 + r^2)}{r^2} - \frac{1}{4} \frac{\tilde{B}^2 c_0}{r^2 (c_0^2 + r^2)}
\eeq
with integration constants $\alpha_1$ and $\alpha_2$. To fix $\alpha_1$ we demand that $\lim_{r\rarrow \infty} Y(r) = 0$ (so that $Y(r)$ does not alter the value of $c_0$), which fixes $\alpha_1=0$. To fix $\alpha_2$, we argue that for sufficiently large $c_0$, the solution for $Y(r)$ must be valid for all $r$. In particular, $Y(r)$ must be finite as $r \rarrow 0$, which means that the coefficient of the $1/r^2$ term must vanish as $r \rarrow 0$, which means we must have
\beq
\alpha_2 =  \frac{1}{2} \, c_0 \left( \partial \phi \right)^2 + \frac{1}{4} c_0 \left( \partial \phi \right)^2 \, \log(c_0^2) + \frac{1}{4} \frac{\tilde{B}^2}{c_0}
\eeq
Inserting this value of $\alpha_2$ into $Y(r)$ and then extracting $c_2$ from the $r \rarrow \infty$ limit, we find
\beq
c_2 = \frac{1}{4} \, c_0 \left( \partial \phi \right)^2 + \frac{1}{4} c_0 \left( \partial \phi \right)^2 \, \log(c_0^2) + \frac{1}{4} \frac{\tilde{B}^2}{c_0}.
\eeq
We can then see from eq. (\ref{omvformula}) that as $c_0 \rarrow \infty$ we indeed have $\Omv \rarrow 0$, for our particular choice of counterterms.

For the $c_0=0$ solutions that we want the finite counterterms, and the $c_0$ terms in $\Omv$, will not contribute. In numerical calculations, however, we necessarily deal with solutions for which $c_0$ is nonzero. We have used the counterterms above for all of our numerical calculations.

Now for $\Ophiv$, which is simpler. We take
\beq
\Ophiv = \, \frac{\delta \Omega}{\delta \phi(\L)} \, = \, - \lim_{\L \rarrow \infty} \left( \frac{\delta S_{reg}}{\delta \phi(\L)} + \sum_i \frac{\delta L_i}{\delta \phi(\L)} \right)  \, = \, \left . \frac{\delta \lag}{\delta \phi'} \right |_{r=\L} \, = \, c.
\eeq
where  in the final equality we have used eq. (\ref{spinningconstantofmotion}) for the constant of motion, $\frac{\delta \lag}{\delta \phi'} = c$. Notice that the counterterms contribute nothing.

Lastly, we want to compute the $U(1)_R$ density $\jtv$. In the SYM theory we have $\jtv = - \frac{d \Omega}{d \mu}$, so in supergravity language we have $\jtv = \frac{d S_{ren}}{d \o}$. We will now restore $\left( \partial \phi \right)^2 = - \o^2$. Let us first compute the contribution from $S_{reg}$, borrowing arguments from ref. \cite{Albash:2007bk}. We start with the regulated action, eq. (\ref{sreg1}), evaluated on a solution. The action is a functional of the fields $y(r)$ and $\phi(t,r)$, and when evaluated on a solution has explicit $\o$ dependence as well as implicit dependence through $y(r)$ and $\phi(t,r)$. We thus use the chain rule:
\beq
\frac{d S_{reg}}{d \o} = - \N \, \int^{\L} dr \, \left [ \frac{\partial \lag}{\partial \o} \, + \, \frac{\partial y}{\partial \o} \frac{\partial \lag}{\partial y} \, + \, \frac{\partial y'}{\partial \o} \frac{\partial \lag}{\partial y'} \, + \, \frac{\partial \phi}{\partial \o} \frac{\partial \lag}{\partial \phi}\, + \, \frac{\partial \phi'}{\partial \o} \frac{\partial \lag}{\partial \phi'} \right]
\eeq
Notice in particular that in the first term the $\frac{\partial}{\partial \o}$ acts only on the explicit $\o$ dependence in $\lag$. We then use the fact that mixed partial derivatives commute to write $\frac{\partial y'}{\partial \o} = \frac{\partial}{\partial r} \frac{\partial}{\partial \o} y$ and similarly for $\phi$, and then integrate by parts to find
\bea
\frac{d S_{reg}}{d \o} & = & - \N \, \int^{\L} dr \, \left [ \frac{\partial \lag}{\partial \o} \, + \, \left( \frac{\partial \lag}{\partial y} \, - \, \frac{\partial}{\partial r} \frac{\partial \lag}{\partial y'} \right) \frac{\partial y}{\partial \o} \, + \, \left( \frac{\partial \lag}{\partial \phi}\, - \, \frac{\partial}{\partial r} \, \frac{\partial \lag}{\partial \phi'} \right) \frac{\partial \phi}{\partial \o} \right] \nonumber \\ & & \,\,\,\,\,\,\, + \left . \frac{\partial y}{\partial \o} \frac{\partial \lag}{\partial y'} \right |_{0}^{\L} + \left . \frac{\partial \phi}{\partial \o} \frac{\partial \lag}{\partial \phi'} \right |_{0}^{\L}
\eea
Clearly the coefficients of the $\frac{\partial y}{\partial \o}$ and $\frac{\partial \phi}{\partial \o}$ terms inside the integral vanish on-shell.

Turning to the boundary terms, we start with those for $y(r)$. Notice first that the contribution from the $r=0$ endpoint vanishes because $\left . \frac{\partial \lag}{\partial y'} \right |_{0} = 0$ due to the $r^3$ factor outside the square root in eq. (\ref{sreg1}). The contribution from the $r=\L$ endpoint also vanishes. To see this we must use the fact that in the SYM theory $m$ and $\mu$ are independent parameters, so that in the supergravity theory $\frac{\partial c_0}{\partial \o}=0$. From the asymptotic form of $y(r)$ in eq. (\ref{spinningyasymptotic}) we can then see that $\frac{\partial y}{\partial \o}$ at $r=\L$ is order $\frac{\log\L}{\L^2}$. That, combined with the fact that $\frac{\partial \lag}{\partial y'}$ in eq. (\ref{dldyprime}) is order one (in the $\L$ counting), indicates that the contribution from the $r=\L$ endpoint is order $\frac{\log\L}{\L^2}$ and hence vanishes as $\L \rarrow \infty$.

For the $\phi$ boundary terms we first identify $\frac{\partial \lag}{\partial \phi'} = c$. Recalling that our ansatz is $\phi(t,r) = \o t + f(r)$, the factor $\frac{\partial \phi}{\partial \o}$ will give us a term which is simply $t$, which vanishes since $ct$ is independent of $r$, so $\left .(ct) \right |_0^{\L} = 0$. As for the $r$ dependence in $\phi$, the contribution from the $r=\L$ endpoint vanishes, as we can see from the asymptotic form of $\phi$ in eq. (\ref{spinningphiasymptotic}): the leading term is order $\frac{1}{\L^2}$ and hence vanishes as $\L \rarrow \infty$. All that remains is the contribution from the $r=0$ endpoint.

We are only interested in $c_0=0$ solutions, however, for which $c=0$ anyway, so that the $\phi$ boundary terms do not contribute at all. In those cases, the only contribution to $\jtv$ comes from the first term under the integral.

As for the counterterms, $L_1$ and $L_3$ contribute nothing, while from the explicit form of $L_2$ in eq. (\ref{explicitcounterterms}) we have that
\beq
\frac{d L_2}{d \o} = + \frac{1}{2} \, \N \, c_0^2 \, \o \, \left( \log (c_0^2) \, - \, \log (\L^2) \, +1 \right) \, + \, O\left( \frac{\log \L}{\L^2} \right) \nonumber
\eeq
Notice that this vanishes for the $c_0=0$ solutions that we want.

To summarize: for the solutions we want, which have $c=0$ and $c_0=0$, and using the solution for $\phi'(r)$ in eq. (\ref{phisolnonzerob}), $\jtv$ is given by
\beq
\jtv = - \N \, \int dr \, \frac{\partial \lag}{\partial \o} = - \N \, \int dr \, r^3 \, \sqrt{1+y'^2} \, \sqrt{\frac{1+\frac{\tilde{B}^2}{(y^2 + r^2)^2}}{1-\o^2 \frac{y^2}{(y^2 + r^2)^2}}} \, \left [ - \, \frac{\o \, y^2}{(y^2 + r^2)^2} \right].
\eeq

Notice also that very similar arguments apply for the magnetization \cite{Albash:2007bk}, given in the SYM theory by $-\frac{d \Omega}{d B}$. The biggest changes are that $\o \rarrow B$, the $\frac{1}{\L^2}$ term in $y(r)$'s asymptotic behavior is relevant to show that the $y$ boundary term vanishes at $r=\L$ rather than the $\frac{\log \L}{\L^2}$ term, and now counterterm $L_3$ contributes rather than $L_2$.

\end{appendix}

\bibliography{spinningd7}
\bibliographystyle{JHEP}

\end{document}